\documentclass[12pt]{article}
\begin{document}
\noindent


\title{An Introduction to Relativistic Quantum Mechanics\\
I. From Relativity to Dirac Equation}
\date{}
\maketitle

\vskip 1.0 truecm

\centerline{M. De Sanctis $^{a,~b}$}
\vskip 1.0 truecm
\noindent
\textit{$^a$ Departamento de F\'isica, Universidad Nacional de Colombia,
Bogot\'a D. C., Colombia.  }

\noindent
\textit{$^b$ INFN sez. di Roma, P.le A. Moro 2, 00185 Roma, Italy. }

\noindent
e-mail :  mdesanctis@unal.edu.co  and   maurizio.desanctis@roma1.infn.it

\vskip 1.0 truecm
\begin{abstract}
\noindent
By using the general concepts of special relativity and the requirements of 
quantum mechanics, Dirac equation is derived and studied. Only elementary knowledge of 
spin and rotations in quantum mechanics and standard handlings of linear algebra
are employed for the development of the present work. 
\vskip 0.25 truecm
\noindent
PACS number(s): 03.30.+p, 03.65.Pm
\end{abstract}

\newpage

\tableofcontents{}
\newpage

\section{Introduction}
\noindent
According to the present knowledge of physics, the ultimate constituents
of matter are \textit{quarks} and \textit{leptons}. Both of them are particles of spin 
$1/2$ that interact by interchanging spin $1$ particles, namely
\textit{photons, gluons,} $W^+,~W^-$ and $Z^0$.
The existence of the Higgs spin $0$ particle is presently under experimental investigation.

\noindent
The issues of relativity  and quantum mechanics, that are strictly necessary 
to understand atomic and subatomic world, have favored the development of local
field theories in which, as we said, the interactions are mediated by the interchange
of the (virtual) integer spin particles mentioned above. 
A general feature of these theories is that, in the field Lagrangian or  Hamiltonian,
the interaction term is \textit{simply} added to   the term that represents
the free motion of the particles.

\noindent
As for  the free term of the matter, spin $1/2$, particles,
it gives rise to the Dirac equation, that represents the relativistic,
quantum mechanical wave equation for these particles.

\noindent 
These arguments  explain  the  great importance of  Dirac equation 
for the study of particle physics at \textit{fundamental} level.
However, it is also strictly necessary to understand   many important aspects 
of atomic physics, nuclear physics and
of the phenomenological models for hadronic particles. 

\vskip 0.5 truecm
\noindent
An introduction to this equation represents the  objective of the 
present work that
is mainly directed to students with good foundations
in nonrelativistic quantum mechanics and some knowledge of special relativity and
classical electrodynamics.

\noindent
We shall not follow the historical development introduced  by Dirac and adopted by
many textbooks. In that case,
the Lorentz transformation (boost) of the Dirac spinors is  performed 
only in a second time, without clarifying 
sufficiently the connection between the mathematics and the physical meaning 
of that transformation.

\noindent
In this paper the Dirac equation will be derived starting from
the basic principles of special relativity and quantum mechanics,
analyzing the transformation properties of the relativistic spinors.

\noindent 
This development will be carried out without entering 
into the mathematical  details of the Lorentz group theory, but keeping 
the discussion at a more \textit{physical} level only using the mathematical
tools of linear vector algebra, as row by column matrix product and vector handling.

\noindent
In our opinion this  introductory approach is highly recommendable in order 
to  stimulate the students
to make independent investigations by using the powerful concept of 
\textit{relativistic covariance}.

\vskip 0.5 truecm
\noindent
In a subsequent work we shall analyze in more detail the properties of Dirac equation 
and derive some  relevant observable effects. 
To that work we shall also defer an introduction to the 
field theory formalism that is needed to give a complete physical description
of subatomic world.

\vskip 0.5 truecm
\noindent
The subjects  of  the present work are examined 
in the following order.

\vskip 0.5 truecm
\noindent
In Subsection 1.1 we give some tedious but necessary  explanations 
about the adopted notation.

\noindent 
In Section 2 we study  some  relevant aspects 
special relativity, focusing our attention on the properties 
of the Lorentz transformations.

\noindent
Their fundamental properties are recalled in Subsection 2.1.

\noindent
We briefly analyze, in Subsection 2.2, classical electrodynamics 
as a \textit{relativistic} fields theory.

\noindent
In Subsection 2.3 we examine the hyperbolic parametrization of the Lorentz transformations,
introducing concepts and techniques that are widely applied in relativistic 
quantum mechanics for the construction of the boost operators.

\noindent
Lorentz transformations in an arbitrary direction are given in subsection 2.4.

\noindent
A very important point of this work is studied in Subsection 2.5, 
where the commutation rules
of the Lorentz boost generators, rotation generators and parity transformation are derived.

\vskip 0.5 truecm
\noindent 
In Section 3 we make use of the concepts of relativity to lay the foundations of
relativistic quantum mechanics.

\noindent
In Subsection 3.1 we discuss,  as an example, the relativistic wave equation 
for a spin 0 particle.

\noindent
In Subsection 3.2 we introduce the (quantum-mechanical) Dirac equation 
for spin $1/2$ particles, starting from 
the commutation rules of the boost generators, rotation generators 
and parity transformation.

\noindent
The properties  of the Dirac Gamma matrices and their different representations
are examined in Subsection 3.3.

\noindent
Some relevant matrix elements of Dirac operators,  as  $\gamma^5$, are studied
in Subsection 3.4.

\noindent 
Finally, plane wave solutions and the corresponding conserved current are found 
and discussed in Subsection 3.5.

\vskip 0.5 truecm
\noindent
The Appendix is devoted to study some useful properties of the Pauli matrices.

\subsection{Notations and Conventions}


\noindent
We suggest the reader to read cursorily this Subsection and to go back to it when 
he finds some difficulty in understanding the other parts of the paper.

\noindent
First of all, the space time position of a particle is denoted as $x^\mu=(x^0,{\bf r})$
with $x^0=ct$ and ${\bf r}= (x^1,~x^2,~x^3)$. 
To avoid confusion, we use this last notation \textit{instead of} the standard one,
that is $(x,~y,~z)$.

\noindent
Greek letters of the ``middle'' part of the alphabet, as $\mu,~\nu,~\rho,~\sigma,...$
running from $0$ to $3$,
are used to denote four-vector components.
On the other hand the letters of the beginning of the Greek alphabet, as 
$\alpha,~\beta,~\delta, ...$ running from $1$ to $3$, denote three-vector components.
This last notation with \textit{upper indices} will be used extensively 
even though the corresponding quantity does not make part of a four-vector.

\noindent
Repeated indices are always \textit{summed}, unless otherwise explicitly stated. 

\noindent
For two three-vectors, say ${\bf a}$ and ${\bf b}$, the scalar product is denoted as
$${\bf a b} = a^\alpha b^\alpha$$ 
If one of the two vectors is a set of the \textit{three} Pauli ($\sigma^\delta$)
or Dirac ($\alpha^\delta$), ($\gamma^\delta$) 
 matrices,  we use the notation
$$({\bf \sigma  a}) = \sigma^\delta a^\delta,~~~~~~ 
 ({\bf \alpha  a}) = \alpha^\delta a^\delta,~~~~~~
 ({\bf \gamma  a}) = \gamma^\delta a^\delta    $$
Furthermore, the notation $\nabla$ collectively indicates the derivatives with respect
to the three components of the position vector ${\bf r}$.

\noindent
Lower indices are only used for four-vectors and denote their \textit{covariant components}
as explained just after eq.(2.2). Invariant product of two four-vectors is introduced
in eq.(2.3).
For the unit vectors we use the standard notation
$${\bf \hat a}={{\bf a}\over |{\bf a}|}$$

\vskip 0.5 cm
\noindent
When a four-vector is used as an \textit{argument} of a field or wave function, the
Lorentz index $\mu,~\nu,~\rho,~\sigma,...$ is dropped and, more simply , we write
$$ A^\mu(x),~~~~~~~~~\psi(x)$$
where $x$ represents collectively all the components of the four-vector $x^\mu$.

\noindent
In order to denote products of matrices and four-vectors, we arrange the components of
a four-vector, say $x^\mu$ in a \textit{column} vector, denoted as $[x]$.
The corresponding \textit{transposed} vector $[x]^T$ is a \textit{row} vector.
Standard Latin letters, without indices,  are used to denote matrices. 
See, for example, eq.(2.6).
We use this notation also for the set of the \textit{four} Dirac matrices $\alpha^\mu$ 
at the end of Subsection 3.2.

\vskip 0.5 truecm
\noindent
Four components Dirac spinors, introduced in Subsection 3.2, are handled according
to the same rules of vector algebra.
They are denoted by a Latin letter \textit{without} parentheses.

\noindent
We recall that the hermitic conjugate of the Dirac spinor $u$
is a row spinor defined as: 
$$u^\dagger={u^*}^T$$ 

\vskip 0.5 truecm
\noindent
For the \textit{commutator} of two matrices (or operators), 
say $Q$, $R$, we use the notation
$$[Q, R]= QR-RQ$$
For the anticommutator we use curly brackets
$$\{Q, R\}= QR+RQ$$

\section{ Relativity}
\vskip 0.5 truecm
\noindent
The principle of relativity, that was found by Galilei and Newton, states that it is 
possible to study physical phenomena from \textit{different} inertial reference frames 
(RF) by means of the \textit{same} physical laws. 
The hypothesis of an \textit{absolute} reference frame is not allowed in physics.

\noindent
Obviously, one has to transform the result of a measurement performed in a reference frame 
to  another reference frame, primarily the measurements of time and space.

\noindent
Requiring the speed of light $c$ to be \textit{independent} of the speed of 
the reference frame, as shown by th Michelson-Morley experiment,
one obtains the \textit{Lorentz transformations}  that represent
the formal foundation of Einstein's special relativity.
The reader can find in ref.[1] a simple and satisfactory development of this point.
\subsection{Fundamental Aspects of Lorentz Transformations}

\noindent
Considering a RF $\cal S'$ moving at velocity $v$ along the $x^1$-axis with respect to
$\cal S$, one has the standard Lorentz transformations

$$x'^0=\gamma(x^0 -{v\over c} x^1)$$
$$x'^1=\gamma(-{v\over c} x^0+ x^1)$$
$$x'^2=x^2$$
$$x'^3=x^3 \eqno(2.1a)$$
where $x^0=ct$, $(x^1, x^2, x^3)={\bf r}$ and 
$\gamma=[1-(v/c)^2]^{-1/2}$. 

\vskip 0.5 truecm
\noindent
A   thorough study of the subject of this Subsection,
that consists in  generalizing the previous equations,
can be found in ref.[2].
In the present paper we highlight some specific aspects that are relevant 
for a quantum-mechanical description of elementary particles.
 
\vskip 0.5 truecm
\noindent
The Lorentz transformations of eq.(2.1a) can be syntetically written as
$$x'^{\mu}=L^{\mu}_{~\nu}(v)x^{\nu} \eqno(2.1b)$$
where the indices $\mu,\nu$ take the values $0,1,2,3$
and $x^\mu$ is denoted as \textit{contravariant} four-vector.

\noindent
By introducing the Minkowsky metric tensor
$$g_{\mu \nu}=g^{\mu \nu}=
\left[ \matrix { 1 ~~~~0  ~~~0 ~~~~0\cr 
                 0 -1 ~~~0 ~~~~0  \cr
                 0 ~~~~0 -1 ~~~~0 \cr 
                 0 ~~~~0 ~~~~0 -1 \cr }
\right]\eqno(2.2)$$
one can construct \textit{covariant} four-vectors
$x_\mu=g_{\mu \nu}x^\nu$ and \textit{invariant}
quantities as \textit{products} of covariant and contravariant four-vectors.
For example, given two contravariant four-vectors, say $s^\mu=(s^0,{\bf s})$ and 
$l^\mu= (l^0,{\bf l})$,
one can construct their covariant counterparts
$s_\mu=(s^0,-{\bf s})$, $l_\mu= (l^0,-{\bf l})$
and the  quantity 
$$s_\mu l^\mu= s^\mu l_\mu = s_\mu g^{\mu \nu} l_\nu = s^\mu g_{\mu \nu}l^\nu =
 s^0 l^0-{\bf s}{\bf l}\eqno(2.3)$$
that is \textit{invariant} under Lorentz transformation:
$$s_\mu l^\mu= s'_\mu l'^\mu \eqno(2.4)$$
\noindent
In particular, the Lorentz transformation of eq.(2.1a) is obtained [1,2] by requiring the 
invariance of the propagation of a spherical light wave, 
that is the invariance of $x^\mu x_\mu=0$.
 
\noindent
The invariance equation (2.4) requires
$$g_{\mu \rho} L^\rho_{~\nu}(v) L^\mu_{~\sigma}(v)= g_{\nu \sigma} \eqno(2.5)$$
In many cases it is very useful to work with standard linear algebra notation.
Furthermore, at pedagogical level, this technique is very useful to introduce
standard handling of Dirac spinors.

\noindent
Identifying a four-vector $x^\mu$ with the column vector $[x]$, the invariant
product of eq.(2.4) is written as
$$s^\mu  g_{\mu \nu}  l^\nu  = [s]^T g [l]\eqno(2.6) $$
where the upper symbol $T$ denotes the operation of transposition. 
By means of this notation, eq.(2.5) reads
$$L(v) g L(v) =g \eqno(2.7)$$
where we have used the important property, directly obtained from eq.(2.1a), that
$L^T(v)=L(v)$.
Also, a covariant four-vector $x_\mu$ is $[x_c]= g[x]$.
Its transformation is
$$[x'_c]=g L(v)[x]=g L(v)gg[x]=g L(v)g[x_c]\eqno(2.8)$$
Let us now multiply eq.(2.7) by $g$ from the right, obtaining

$$L(v) g L(v)g =1 \eqno(2.9)$$
In consequence
$$gL(v)g=L^{-1}(v)\eqno(2.10)$$
it means that the covariant four-vectors, look at eq.(2.8)!, transform with the 
\textit{inverse} Lorentz transformations.
By means of direct calculation or by using the principle of relativity one finds that
$$L^{-1}(v)=L(-v)\eqno(2.11)$$

\vskip 0.5 truecm
\noindent
We recall some relevant physical quantities that are represented by
(i.e. transform as) a four-vector.
As previously discussed,
we have the four-position (in time and space) of a particle denoted by $x^\mu$.

\vskip 0.5 truecm
\noindent
We now define the four-vector that represents the energy and momentum of a particle. 

\noindent
Previously, we introduce the (invariant) rest mass of the particle.
In the present work this quantity will be simply denoted as the \textit{mass} $m$.
We shall \textit{never} make use of the so-called relativistic mass.

\noindent
We also define the differential of the proper (invariant) time as
$$d\tau={1\over c}[dx_\mu dx^\mu]^{1/2}=
\left[(dt)^2-{1\over c^2}(d{\bf r})^2\right]^{1/2}=$$
$$=dt \left[1-{({{\bf v}\over c})}^2 \right]^{1/2}=
{d t\over \gamma} \eqno(2.12)$$
where the velocity 
$${\bf v}={d {\bf r}\over dt}$$ 
represents the standard physical velocity 
of the particle measured by an observer in a given reference frame. 
Furthermore, the factor $\gamma$ is a function of that velocity, 
of the form: 
$$\gamma=\left[1-{({{\bf v}\over c})}^2\right]^{-1/2}$$

\noindent
The energy-momentum four-vector is obtained differentiating the four-position 
with respect to the proper  time 
and multiplying the result by the mass $m$. One has
$$p^\mu= ({E\over c},{\bf p})=  m{{d x^\mu}\over d\tau}=
(mc\gamma, m{\bf v}\gamma) \eqno(2.13)$$
In  previous equation, $E$ represents the energy of the particle and
${\bf p}$ its three-momentum.
More explicitly, the energy is 
$$E=mc^2\gamma$$

\noindent
For small values of the velocity $|{\bf v}|<< c$ one recovers 
the nonrelativistic limit, that is

$$E \simeq mc^2 +{1\over 2} m{\bf v}^2+...\eqno(2.14a)$$
$${\bf p} \simeq m{\bf v}+...\eqno(2.14b)$$ 
Note that the energy and momentum of a particle belong to the four-vector of eq.(2.13).
In consequence, energy and momentum conservation can be written 
in a \textit{manifestly covariant} form.
For example, in a collision process in which one has a transition from  
an initial state (I) with $N_I$ particles, to a final state (F) with $N_F$ particles,
the total energy and momentum conservation is written
by means of the following four-vector equality

$$\sum_{i=1}^{N_I} p_i^\mu(I)= \sum_{i=1}^{N_F} p_i^\mu(F)\eqno(2.15)$$
that holds in any reference frame.
A complete discussion of the physical consequences of that equation 
and  related matter is given in ref.[3].
Only recall that,
at variance with nonrelativistic mechanics,
mass is \textit{not conserved}. 
In general, mass-energy transformations  are represented by processes 
of creation and destruction of particles.
As a special case, a scattering reaction is defined \textit{elastic},
if all the particles of the final state
remain \textit{the same} (obviously, with the same mass) 
as those of the initial state. 

\vskip 0.5 truecm
\noindent
Four-momentum conservation of eq.(2.15) is a very simple example. 
In general, a physical law written in 
a \textit{manifestly covariant} form  automatically fulfills the principle of relativity
introduced at the beginning of this section.

\noindent
A physical law is written in a \textit{manifestly covariant} form
when it is written as
an equality between two relativistic tensors of the same rank: 
two Lorentz invariants (scalars), two four-vectors, etc..

\vskip 0.5 truecm
\noindent
Going back to eq.(2.13) one can construct the following invariant
$$p^\mu p_\mu=\left({E\over c}\right)^2 -{\bf p}^2 =(mc)^2\eqno(2.16)$$
The second equality is obtained in the easiest way by calculating
the invariant  in
the \textit{rest frame} of the particle, where $p^\mu=(mc,{\bf 0})$.

\noindent
From the previous equation one can construct the Hamiltonian of a particle, that is 
the energy written as function of the momentum
$$E=[({\bf p} c)^2 + (mc^2)^2]^{1/2}\eqno(2.17)$$
that in the nonrelativistic limit reduces to

$$E\simeq mc^2+ {{\bf p}^2 \over{2m}} +...$$
Note that in eq.(2.17) we have  taken only the \textit{positive} value of the square rooth. 
This choice  is perfectly legitimate in a classical context, where the energy 
changes its value in a continous way. On the other hand negative energy solutions 
cannot be discarded when considering quantum-mechanical equations.

\vskip 0.5 truecm
\noindent
From eqs.(2.13) and (2.17), the velocity of a particle is
$${\bf v}={{{\bf p}\over E}}~,~~~~~~|{\bf v}|\leq c$$
In the second relation, the equality is satified by \textit{massless} particles.
The constraint on velocity has a more general validity,
as we shall see  when revising electromagnetism:
\textit{a physical particle } cannot have a velocity
greater than the speed of light $c$.

\noindent 
In this concern we observe that
the word \textit{information} has been also
used but it should be clarified   what does the word 
\textit{information} really mean.
Furthermore, the concepts and the requirements of quantum mechanics about
physical states and measurements have not been  taken into account when introducing
that constraint.

\vskip 0.5 truecm
\noindent
For the study of both classical and quantum-mechanical (field) theories 
it is very important
to determine the transformation properties of the derivative operator
$${\partial \over {\partial x^\mu}}=\left(
{1\over c}{{\partial \over\partial t}},{{\partial \over\partial{\bf r}}}
\right)=
\left(
{1\over c}{{\partial \over\partial t}}, \nabla
\right)$$
The reader is suggested to derive them by using directly the chain rule.
We propose here a simpler proof.
Let us consider the invariant $x^\nu x_\nu=(x^0)^2-{\bf r}^2$ 
and apply to it the derivative operator.
One has
$${\partial \over {\partial x^\mu}} x_\nu x^\nu =2x_\mu=2(x^0,-{\bf r})\eqno(2.18a)$$ 
That is, the derivative with respect to the contravariant components gives,
\textit{and transforms as }, a 
covariant four-vector ($2x_\mu$ in the previous equation). 
Conversely, the derivative with respect to the covariant components 
\textit{transforms as} a 
contravariant four-vector: 
$${\partial \over {\partial x_\mu}} x_\nu x^\nu =2x^\mu=2(x^0,{\bf r})\eqno(2.18b)$$ 
For this reason the following notation is introduced
$${\partial \over {\partial x^\mu}} = \partial_\mu \eqno(2.19a)$$
and
$${\partial \over {\partial x_\mu}} = \partial^\mu \eqno(2.19b)$$
Straightforwardly one verifies that

$${\partial \over {\partial x^\mu}} {\partial \over {\partial x_\mu}}=
 \partial_\mu \partial^\mu =
{1\over c^2}{{\partial^2 \over\partial t^2}}- \nabla^2
\eqno(2.20)$$
is an \textit{invariant} operator.
\subsection{Electromagnetism and Relativity}

\noindent
The elements that have been developed in the preceding Subsection
will help us to understand the relativistic  properties of classical electromagnetism.

\noindent
In summary, electromagnetism is a \textit{local} theory in which the interaction 
between charged particles is carried by the electromagnetic \textit{field}, 
at light speed $c$.
A complete analysis of this theory can be found, for example, in refs.[2,4].

\vskip 0.5 truecm
\noindent
With respect to interaction propagation, the reader should realize that
Newton's theory of gravitational interaction is not compatible
with special relativity. In fact the gravitational potential energy
$$V_g=-{{G m_1 m_2}\over r}$$
depends \textit{instantaneously} on the distance $r$ between the two bodies.
If one body, say the $\# 1$, changes its position or state, 
the potential energy, and in consequence, \textit{the force} 
felt by the body $\# 2$ changes at the \textit{same instant},
implying a transmission of the interaction at \textit{infinite} velocity.

\noindent
Note that,
on the other  hand, the expression of
Coulomb potential energy, that is formally analogous to the Newton's
gravitational one, holds exactly \textit{exclusively in the static case}.
According to classical electromagnetism,  
if the interacting particles are in  motion, it represents
\textit{only  approximatively} their interaction. 
This approximation is considered good if their relative velocity is
$$|{\bf v}|<< c$$

\vskip 0.5 truecm
\noindent
The fundamental quantity of electromagnetism 
is the vector potential \textit{field} $A^\mu =(A^0,{\bf A})$. 

\noindent
A field is, by definition, a function of the time-space position $x^\nu$. 
As done in most texbooks,
in the following we shall drop the index $\nu$ of the argument, 
simply writing $A^\mu= A^\mu(x)$. 

\noindent
Synthetically, we recall that the Maxwell equations have the  form

$$\partial_\nu \partial^\nu A^\mu={{4\pi}\over c}j^\mu \eqno(2.21)$$
with the Lorentz invariant Gauge condition
$$\partial_\mu A^\mu=0 \eqno(2.22)$$
where we have introduced the \textit{current density}
$$j^\mu =(c\rho(x),{\bf j}(x))\eqno(2.23)$$
Applying the derivative operator $\partial_\mu$ to eq.(2.21) and using eq.(2.22),
one finds the current conservation equation, that is
$$\partial_\mu j^\mu= 
{{\partial}\over{\partial t}}\rho(x)+
{{\partial}\over{\partial {\bf r}}}{\bf j}(x)
= 0 \eqno(2.24)$$
All the equations written above are \textit{manifestly covariant} and the 
Lorentz transformations can be easily performed.
If a solution of eqs.(2.21) and (2.22) is found in a reference frame $\cal S$,
it is not necessary to solve the equations in the reference frame $\cal S'$, 
but simply one can transform  the electromagnetic field:
$$A'^\mu(x')=L^\mu_{~\nu}(v)A^\nu(x(x'))\eqno(2.25)$$  
In more detail, one has

\noindent
(i) to tranform the field $A^\mu$, mixing its components by means of
$L^\mu_{~\nu}(v)$, that is the first factor of the previous equation, but also 

\noindent
(ii) to express the argument $x$ of the frame $\cal S$ as a function of $x'$ 
measured in $\cal S'$, that is, recalling eq.(2.11) 
$$x^\nu=L^\nu_{~\rho}(-v)x'^\rho$$
We briefly define the last operation as \textit{argument re-expression}.

\vskip 0.5 truecm
\noindent
The reader should note that such \textit{double} transformation occours in the same way
when  a rotation is performed. 
In this case the space components ${\bf A}$ are mixed by the rotation matrix 
(for this reason the electromagnetic field is defined as a \textit{vector} field)
and the argument ${\bf r}$ must be expressed in terms of ${\bf r}'$
by means of the inverse rotation matrix.
Under rotation, in the time component $A^0$, one only has the argument 
re-expression of ${\bf r}$.

\vskip 0.5 truecm
\noindent
In principle it is possible to construct a \textit{scalar} field theory
(even though  there is no evidence of such theories  at macroscopic level).
In this case the field is represented by a one-component function $\phi(x)$.
Both the Lorentz transformation and the rotations only affect the
argument $x$ in the same way as before, but no mixing can occur for the 
\textit{single} 
component function $\phi$. 
One has only to perform the \textit{argument re-expression}.

\vskip 0.5 truecm
\noindent
We shall now explain with a physical relevant example the use of the transformation
(2.25) for the electromagnetic field.

\noindent
Let us consider a  charged particle moving with velocity $u$ along the $x^1$-axis.
What is the field produced by this particle ?

\noindent
We introduce a reference frame $\cal S$ in which the particle is at rest, while
the observer in $\cal S'$ sees the particle moving with velocity $u$ along $x^1$.
The velocity of $\cal S'$  with respect to $\cal S$ is $v=-u$.
The field in $\cal S$ is purely electrostatic, that is
$$A^0=A^0(ct,{\bf r})={q\over{|{\bf r}|}}\eqno(2.26a)$$
$${\bf A}={\bf A}(ct,{\bf r})={\bf 0} \eqno(2.26b)$$
where $q$ represents the charge of the particle.
We find $A'^\mu$ by means of eq.(2.25). First, one has
$$A'^0=\gamma A^0 \eqno(2.27a)$$
$$A'^1={u\over c} \gamma A^0 \eqno(2.27b)$$
$$A'^\alpha=A^\alpha=0 \eqno(2.27c)$$
with $\alpha=2,3$ and $\gamma=[1-(u/c)^2]^{-1/2}$. 

\noindent
Now we express $|{\bf r}|$ in terms of $(ct',{\bf r}')$, 
that is we perform  the
\textit{argument re-expression}.

\noindent
By means of eqs.(2.1) and (2.11) one has
$$x^1=\gamma(-ut'+x'^1)$$
$$x^\alpha=x'^\alpha$$
so that
$$|{\bf r}|=[\gamma^2 (-ut'+x'^1)^2 +(x'^2)^2+(x'^3)^2]^{1/2} \eqno(2.28)$$
By means of the previous equation the final expression for the field of 

\noindent
eqs.(2.27a,b) is
$$A'^0(ct',{\bf r}')=
q \gamma[\gamma^2 (-ut'+x'^1)^2 +(x'^2)^2+(x'^3)^2]^{-1/2}\eqno(2.29a)$$ 
$$A'^1(ct',{\bf r}')={u\over c}A'^0(ct',{\bf r}')\eqno(2.29b)$$
This example has been chosen to explain the procedure for transforming a field function.

\noindent
The field of eqs.(2.29a,b) can be directly derived in the frame
$\cal S'$ by solving the Maxwell equations (2.21),(2.22) as done in refs.[2,4]. 
The technique of the 
Li\'enard Wiechert potentials can be used. But, as the reader should check, 
much more mathematical  efforts are required.
\subsection{The Hyperbolic Parametrization of the Lorentz Transformations}

\noindent
Going back to the Lorentz transformations of eq.(2.1) we note that the coefficients
of the  of the transformation matrix $L(v)$ are:
\vskip 0.5 truecm

$$\gamma~,~~~~~~~~~~~-{v\over c}\gamma$$

\vskip 0.5 truecm
\noindent 
We square both terms (the minus sign disappears in the second one)
and take the difference, obtaining
$$\gamma^2 -\left({v\over c}\gamma\right)^2=1$$
Recalling that the hyperbolic functions satisfy the relation
$$ ch^2\omega-sh^2\omega=1$$
one can choose the following parametrization
$$\gamma=ch~ \omega ~,~~~ {v\over c}\gamma=sh~\omega\eqno(2.30a)$$
with
$${v\over c}=th~\omega \eqno(2.30b)$$
that connects the \textit{hyperbolic} parameter $\omega$ with the standard velocity $v$.
We now show the reason why the parametrization $L(\omega)$
is very useful for the following developments.

\noindent
Let us consider two subsequent Lorentz transformations
along the $x^1$ axis with hyperbolic parameters 
$\eta$ and $\xi$. The total transformation is given by the following product of
Lorentz transformations, that, by using the vector algebra notation, 
is written in the form
$$[x']=L(\eta) L(\xi)[x]\eqno(2.31)$$
The reader can calculate  explicitly $L(\eta) L(\xi)$
by means of standard rules for row by column matrix product, then recalling
$$sh( \eta +\xi)= sh~\eta ~ch~\xi+ sh~\xi ~ch~\eta$$
$$ch( \eta +\xi)= ch~\eta ~ch~\xi+ sh~\xi ~sh~\eta$$
one finds
$$L(\eta) L(\xi)=L(\eta+\xi)\eqno(2.32)$$
Note that the previous result strictly depends on the chosen hyperbolic parametrization.
Due to the relativistic \textit{nonlinear} composition of velocities, 
considering two subsequent Lorentz transformations, with
$v/c= th~\eta$ and $w/c=th~\xi$,
one has, in contrast to eq.(2.32),
$$L(v) L(w)\neq L(v+w)\eqno(2.33)$$
On the other hand, the composition of two Lorentz transformations with hyperbolic
parametrization, as given in eq.(2.32), has the same form  as the composition 
of two rotations around the same axis.

\noindent 
Eq.(2.32) is the clue for the following development.

\vskip 0.5 truecm
\noindent
By means of hyperbolic parametrization,
we can now turn  to express a finite Lorentz 
transformation in terms of the corresponding infinitesimal transformation.
  
\noindent
Let us consider the case of \textit{small} velocity, that is
$v/c << 1$  or equivalently,  for the hyperbolic parameter, 
$ \omega \simeq 0$ (see eq.(2.30b)).

\noindent
In particular, at first order in $\omega$ or in $ v/ c$,
one has
$$~ch~\omega \simeq 1~,~~sh~\omega \simeq \omega \simeq {v\over c}$$
and, in consequence
$$L(\omega)\simeq  1 +\omega K^1 \eqno(2.34)$$
where $ 1$  and $K^1$ respectively represent the identity matrix and the 
\textit{generator} of the Lorentz transformation  matrix along the $x^1$ axis.
This second term is usually called \textit{boost generator}.

\noindent 
Explictly, the matrix $K^1$ is easily obtained considering  eqs.(2.1), (2.31a) 
and the above Taylor expansions of the hyperbolic functions. It has the form

$$K^1= \left[ \matrix {~~  0     -1 ~~~~0 ~~~~0 \cr 
                          -1  ~~~~0 ~~~~0 ~~~~0 \cr
                       ~~~ 0  ~~~~0 ~~~~0 ~~~~0 \cr 
                       ~~~ 0  ~~~~0 ~~~~0 ~~~~0 \cr }
\right]=
-\left[ \matrix{ \sigma^1 ~~ 0\cr
                  0   ~~~   0  }   \right]\eqno(2.35)$$
The second expression in the previous equation is given for pedagogical reasons,
that is to familiarize the reader with \textit{block} matrices.

\noindent
In fact, the $4\times 4$ matrix $K^1$ is written as a \textit{block} matrix, 
in which each block is represented by a $2\times 2$ matrix. 
In particular the upper left block is the Pauli matrix 
$\sigma^1$, in the other ``$0$'' blocks the four entries of each block are all vanishing.
The properties of the Pauli matrices are studied in the Appendix. 
For their definition see eq.(A.1).

\noindent
The reader should note the following two points:

\noindent
(i) there is no direct connection between $\sigma^1$ of the previous equations
and the quantum mechanical spin operator,

\noindent 
(ii) as for the row by column product of a block matrix, the same rules 
of standard matrices must be used. 

\vskip 0.5 truecm
\noindent
In order to reconstuct the finite boost $L(\omega)$, ($\omega$ finite),
we apply $N$ times, with  $N \rightarrow \infty$, the infinitesimal
transformation of eq.(2.34).

\noindent
The \textit{linear} boost composition law of eq.(2.32) allows to derive 
the following equation

$$L(\omega)=\lim_{N\rightarrow \infty}\left( 1 
+{\omega \over N} K^1 \right)^N 
=\exp(\omega K^1)\eqno(2.36a)$$
$$= 1+ (ch ~\omega -1)\left[ \matrix{ 1 ~~~ 0\cr
                  0   ~~~   0  }   \right]
+sh~\omega K^1\eqno(2.36b)$$
The second equality of eq.(2.36a)is obtained by comparing the series expansion
in powers of $\omega$ of the exponential, with 
$\left( 1 + {\omega \over N} K^1 \right)^N$ for  ${N\rightarrow \infty}~$.

\noindent
Eq.(2.36b) is derived working on that series expansion. 
One has the following rules for the powers of $K^1$
$$(K^1)^0= 1~,~~~~~~
  (K^1)^{2n}=  \left[ \matrix{ 1 ~~~ 0\cr
                  0   ~~~   0  }   \right]~,~~~~~~
  (K^1)^{2n+1}= K^1$$
that, for example, can be derived from the corresponding properties of $\sigma^1$
by means of eq.(A.4).

\noindent
The coefficients that multiply $(K^1)^{2n}$  and  
$(K^1)^{2n+1}$   can be \textit{summed up},
giving $ch~\omega-1$ and $sh~\omega$,
respectively. One can straightforwardly check that  eq.(2.36b)
is equal to  eq.(2.1) with the hyperbolic parametrization of eq.(2.30a).
We have developed in some detail this example as a guide to construct the finite boost
transformations for Dirac spinors in eqs.(3.17a,b).

\vskip 0.5 truecm
\noindent
We remind the reader that all the relevant properties of the boost transformation
are contained in the infinitesimal form given in eq.(2.34) with the matrix boost
generator of eq.(2.35). The finite expression of the boost is obtained by means of 
a standard mathematical procedure that does not add new physical information.

\subsection{Lorentz Transformations in an Arbitrary Direction }

\noindent
In the previous developments we have considered Lorentz transformations along 
the $x^1$-axis. The transformations along the  $x^2$-  and $x^3$-axis
are directly obtained
interchanging the spatial variables. In this way  (as done in ref.[4]) one
obtains a sufficiently general  treatment of relativistic problems.
For completeness and to help the reader with the analysis of some textbooks
(as for example ref.[2]) and research articles,
we now study Lorentz transformations with an arbitrary boost velocity ${\bf v}$ direction.
The comprehension of the other Sections of this work does not depend on this point.
In consequence, the reader (if not interested)  can go directly to eq.(2.41).

\noindent
For definiteness we consider the time-space four-vector $x^\mu=(x^0,{\bf r}) $, 
but the results hold
for \textit{any} four-vector.

\noindent
The transformation equation (2.1) can be generalized in the following way

$$x'^{~0}=\gamma(x^0 -{{\bf r v}\over c}) \eqno(2.37a)$$
$${\bf r '\hat v}=\gamma(-{v\over c} x^0+  {\bf r \hat v}) \eqno(2.37b)  $$
$${\bf r'}_{\perp}= {\bf r}_{\perp}\eqno(2.37c)    $$
where the unit vector ${\bf \hat v}$ has been introduced so that 
${\bf v}=v{\bf \hat v}$  with $v>0$
and the notation ${\bf r}_\perp $ denotes the spatial components of ${\bf r}$
perpendicular to ${\bf v}$.

\noindent
Eq.(2.37a) directly represents the Lorentz transformation of 
the time component of a four-vector
for an arbitrary direction of the boost velocity. 

\noindent
Some handling is necessary for the spatial components of the four-vector.
Starting from eqs.(2.37b,c) we now develop
the transformation for ${\bf r}$. One can parametrize this transformation according
to the following hypothesis
$${\bf r'}={\bf r}+ f(v){1\over c^2}({\bf r v}){\bf v}+
g(v){1\over c}x^0{\bf v}\eqno(2.38)$$ 
Note that it correctly reduces to the identity when $v=0$ and automatically gives 
eq.(2.37c) for the perpendicular components of the four-vector.

\noindent
Multiplying the previous equation by ${\bf \hat v}$ and comparing with eq(2.37b)
one finds

$$g(v)= -\gamma \eqno(2.39a)$$
$$f(v)={{\gamma-1}\over ({v\over c})^2}={\gamma^2\over{\gamma+1}}\eqno(2.39b)$$
where the last expression of eq.(2.39b) is obtained by using the standard definition 
of the factor $\gamma$.

\noindent
Analogously to eq.(2.36a), the Lorentz transformation for the four-vector given by
eqs.(2.37a) and (2.38) can be written in exponential form, as
$$L(\omega {\bf \hat v})=\exp(\omega {\bf \hat v K})\eqno(2.40)$$
with the same connection between $\omega$ and $ v$ as in eq.(2.30b). 
The matrices of the boost generator 
${\bf K}=(K^1,K^2,K^3)$
are defined as

$$K^1= \left[ \matrix {~~  0     -1 ~~~~0 ~~~~0 \cr 
                          -1  ~~~~0 ~~~~0 ~~~~0 \cr
                       ~~~ 0  ~~~~0 ~~~~0 ~~~~0 \cr 
                       ~~~ 0  ~~~~0 ~~~~0 ~~~~0 \cr }\right]
~
  K^2= \left[ \matrix {~~  0  ~~~~0  ~ -1 ~~~~0 \cr
                       ~~~ 0  ~~~~0 ~~~~0 ~~~~0 \cr 
                          -1  ~~~~0 ~~~~0 ~~~~0 \cr
                       ~~~ 0  ~~~~0 ~~~~0 ~~~~0 \cr }\right]
~
  K^3= \left[ \matrix {~~  0  ~~~~0 ~~~~0   ~ -1 \cr
                       ~~~ 0  ~~~~0 ~~~~0 ~~~~0 \cr 
                       ~~~ 0  ~~~~0 ~~~~0 ~~~~0 \cr
                          -1  ~~~~0 ~~~~0 ~~~~0 \cr }\right]\eqno(2.41)$$
Note that $K^1$ had been already derived in eq.(2.35).
Furthermore, $K^2$ and $K^3$ can be directly 
obtained performing the Lorentz transformation analogously to eq.(2.1) but along the 
$x^2$-  and $x^3$-axis and repeating the procedure that leads to eq.(2.35). 
\subsection{The Commutation Rules of the Boost Generators }

\noindent
The most important property of the boost generators or, 
more precisely, of the matrices ${\bf K}$ given in eq.(2.41), is represented
by their \textit{commutation rules}.
Let us consider an illustrative example. For generality, we shall denote 
the Lorentz transformation as \textit{boost}, using the symbol $B$.

\noindent
In a \textit{first step},
we perform a  boost along  $x^2$ with a \textit{small} velocity. At first order in the
hyperbolic parameter $\omega^2$ one has
$$ B^2\simeq  1 +\omega^2 K^2$$
Analogously, in a \textit{second step}, we make a boost along $x^1$, with hyperbolic
parameter $\omega^1$, that is 
$$ B^1\simeq  1 +\omega^1 K^1$$
The total boost, up to order $ \omega^1 \omega^2 $, is 
$$B^{12}=B^1 B^2 \simeq  1 +\omega^1 K^1+\omega^2 K^2
+\omega^1\omega^2 K^1 K^2\eqno(2.42a)$$
Note the important property that the product of two boosts is a Lorentz boost 
because it satisfies eq.(2.7), as it can be directly verified.

\noindent
We now repeat the previous procedure \textit{inverting} the order of the two boosts, 
obtaining
$$B^{21}=B^2 B^1 \simeq  1 +\omega^1 K^1+\omega^2 K^2
+\omega^1\omega^2 K^2 K^1\eqno(2.42b)$$
What is the difference between the two procedures ? Subtraction of eqs.(2.42a,b) gives
$$B^{12}-B^{21}\simeq \omega^1\omega^2 [K^1,K^2]\eqno(2.43)$$
where the standard notation for the commutator of the matrices  $K^1$ and $K^2$ has been
introduced. Explicit calculation gives

$$  [K^1,K^2]=      \left[ \matrix { ~~~ 0  ~~~~0 ~~~~0 ~~~~0 ~ \cr 
                                     ~~  0  ~~~~0 ~~~~1 ~~~~0   \cr
                                     ~~~ 0   ~~ {-1}  ~~~~0 ~~~~0 ~ \cr 
                                     ~~~ 0  ~~~~0 ~~~~0 ~~~~0 ~ \cr }
 \right]\eqno(2.44)$$
At this point two (connected) questions are in order. What is the meaning of the
\textit{noncommutativity} of the boost generators? Which physical quantity is represented
by the  commutator of the last equation?

\noindent
To answer these questions it is necessary to recall some properties of the 
\textit{rotations}.

\noindent
They are initially defined in the three dimensional space. Let us rotate the vector
${\bf r}$ counterclockwise, around the $x^3$ axis, of the angle $\theta^3$. 
For a small angle, at first order in $\theta^3$, one obtains 
the rotated vector 
$${\bf r'} \simeq {\bf r} + \theta^3 {\bf \hat k} \times {\bf r}\eqno(2.45a)$$
where ${\bf \hat k}$ represents the unit vector of the $x^3$ axis. 
One can put ${\bf r}$ and ${\bf r'}$ in the three component column vectors
$[{\bf r}]$ and $[{\bf r'}]$ so that the previous equation  can be written with 
the vector algebra notation as
$$[{\bf r'}] \simeq 
\left(  1 + \theta^3 s^3 \right)[{\bf r}]\eqno(2.45b)$$
where $s^3$ (see the next equation) represents the  three-dimensional generator 
matrix of the rotations around the axis $x^3$. 
The same procedure can be repeated for the rotations around 
the axes $x^1$ and $x^2$. The generator matrices are
$$ s^1= \left[ \matrix {  ~~~~0 ~~~~0 ~~~~0 ~ \cr 
                          ~~~ 0 ~~~~0  ~~ {-1} \cr 
                          ~~~~0 ~~~~1 ~~~~0 ~ \cr }\right],~
   s^2= \left[ \matrix {  ~~~~0 ~~~~0 ~~~~1 ~ \cr 
                          ~~~~0 ~~~~0 ~~~~0 ~ \cr 
                        ~~ {-1} ~~~~0 ~~~~0 ~ \cr }\right],~
   s^3= \left[ \matrix {  ~~~~0 ~~ {-1}  ~~~~0 \cr
                          ~~~~1 ~~~~0 ~~~~0  \cr
                          ~~~~0 ~~~~0 ~~~~0}\right]
\eqno(2.46)$$
As it is well known the (previous) rotation generator matrices do not commute:
$$[s^{\alpha},s^{\beta}]=\epsilon^{\alpha \beta \delta}s^{\delta}\eqno(2.47)$$
where we have introduced the Levi-Civita antisymmetric tensor 
$\epsilon^{\alpha \beta \delta}$.

\noindent
As for the \textit{noncommutativity},
this situation is partially similar to the case of the boost generators shown in 
eq.(2.43) but, for the rotations, eq.(2.47) shows that, given two generator matrices,
their commutator is proportional to the \textit{third} matrix, while we have not yet
identified the physical meaning of the matrix in the \textit{r.h.s.} of eq.(2.44).

\vskip 0.5 truecm
\noindent
Pay attention ! In quantum mechanics, from eqs.(2.46), (2.47)
we can introduce the \textit{spin 1} operators as
$$j_1^\alpha=-i \hbar s^\alpha$$
satisfying the standard angular momentum commutation rules.
Our $s^\alpha$ do not \textit{directly} represent the three spin operators.

\vskip 0.5 truecm
\noindent
It is very important to note that the physical laws must be \textit{invariant}
under rotations. To make physics we assume that space is \textit{isotropic}.
As the hypothesis of an absolute reference frame must be refused,
in the same way 
the idea of a \textit{preferencial} direction in the space is not allowed by the 
conceptual foundations of physics.

\noindent
Obviously, rotational invariance must be  compatible with relativity. 
This fact is immediately evident recalling that the rotations \textit{mix}
the spatial components of a vector without changing the scalar product of two
three-vectors, say ${\bf a}$ and ${\bf b}$:
${\bf a'b'}={\bf a b}$. 

\noindent
The  time components of the corresponding four-vectors
also remain unaltered: $a'^{~0}=a^0$ and $b'^{~0}=b^0$. Consider, as two relevant examples,
the time and energy that represent the zero components of the position and momentum
four-vectors, respectively.

\noindent
It means that rotations satisfy
the invariance equation (2.3) and, in consequence, they are fully compatible
with relativity. 
In terms of $4\times 4$ matrices, eq.(2.45b) is generalized  as 
$$[x'] \simeq 
\left(  1 + \theta^3 S^3 \right)[x]\eqno(2.48)$$
The $4\times 4$ generator matrices are defined in terms of $3 \times 3$ $s^\alpha$ as

$$ S^{\alpha}= \left[ \matrix {   ~~~0 | ~~~0 ~~~0  ~~~0 \cr
                                      ~~~  ------ \cr
                                 ~~~~0 | ~~~~~ ~~~~~  ~~~~~ \cr
                                  ~~~0 | ~~~~~~~  s^{\alpha} ~~~~ \cr
                                 ~~~~0 | ~~~~~ ~~~~~  ~~~~~ \cr
}\right]\eqno(2.49)$$
As it will be written in eq.(2.50b), the matrices $S^{\alpha}$ obviously
satisfy the same commutation rules of eq.(2.47).

\vskip 0.5 truecm
\noindent
We can now verify that the \textit{r.h.s.} of eq.(2.44) represents $~-S^3$. 

\noindent
In general one has the following commutation rules
$$[K^{\alpha}, K^{\beta}]=-\epsilon^{\alpha \beta \delta} S^\delta \eqno(2.50a)$$ 
For completeness, we also give 
$$[S^{\alpha}, S^{\beta}]= \epsilon^{\alpha \beta \delta} S^\delta \eqno(2.50b)$$ 
and
$$[S^{\alpha}, K^{\beta}]= \epsilon^{\alpha \beta \delta} K^\delta \eqno(2.50c)$$ 
where the last equation means that the boost generator ${\bf K}$ transforms as a
vector under rotations. 

\vskip 0.5 truecm
\noindent
As for the derivation of the Dirac equation that will be performed in the next section,
we anticipate here that a set of $K^{\alpha}$ and $S^{\alpha}$ matrices 
(different from eqs.(2.41) and (2.46),(2.49)) will be found, that satisfy the 
\textit{same} commutation rules of eqs.(2.50a-c). 
In mathematical terms, these new matrices are a different \textit{representation}
of the Lorentz group, allowing to satisfy in this way the relativistic invariance 
of the theory.

\vskip 0.5 truecm
\noindent
For the study of the Dirac equation, it is also necessary to introduce another 
invariance property related to a new, discrete,  \textit{space-time} transformation.
It is the \textit{parity transformation}, or \textit{spatial inversion}, 
that changes the position three-vector ${\bf r}$ into $-{\bf r}$, 
leaving the time component unaltered. 
This definition shows that spatial inversion does not change the invariant product 
of two four-vectors and, in consequence, is \textit{compatible} with relativity.

\noindent
Parity is a \textit{discrete} transformation that does not depend on any parameter.
On the other hand, recall that  rotations are 
\textit{continous} tranformations, that \textit{continously} depend on the rotation
angle. 
Obviously, spatial inversion cannot be accomplished by means of  rotations.

\vskip 0.5 truecm
\noindent
Note that, under parity transformation, ordinary, or \textit{polar} vectors, 
as for example the momentum ${\bf p}$, 
do change sign in the same way  as the position ${\bf r}$,
while the \textit{axial} vectors, as for example the orbital angular momentum 
${\bf l}={\bf r}\times {\bf p}$,
do not change sign. 
On the other hand they transform in standard way under rotations.

\vskip 0.5 truecm
\noindent
Using the  definition given above, parity transformation on the space-time position,
 $$[x']=\Pi [x]$$
is accomplished by means of the diagonal Minkowsky matrix.
We can write
 $$\Pi= g$$
that holds for the spatial inversion of \textit{all} the four-vectors.

\noindent
From the previous definition, one can easily verify
the following \textit{anticommutation} rule with the boost generators
$$\{ \Pi, K^{\alpha} \}=0 \eqno(2.51a)$$
or equivalently
$$\Pi K^{\alpha} \Pi= -K^{\alpha} \eqno(2.51b)$$
where we have used the standard property $\Pi^2= 1$.

\noindent
Furthermore
$$[\Pi, S^{\alpha}]=0\eqno(2.52a)$$
or equivalently
$$\Pi S^{\alpha} \Pi= S^{\alpha}\eqno(2.52b)$$
It shows that  the rotation generators do not change sign under spatial inversion,
that is they behave as an axial-vector.

\noindent
The \textit{determinant} of Lorentz boost and rotations is 
equal to $+1$, while for spatial inversion it is $-1$.

\vskip 0.5 truecm
\noindent
Note that eqs.(2.51a)-(2.52b) represent \textit{general} properties of the parity
transformation that do not depend on the tensor to which it is applied. 
They are derived, and hold, in the case of four-vectors, but they are also
assumed to hold for the Dirac spinors. 
But, in this case, the following critical discussion is necessary.

\vskip 0.5 truecm
\noindent
In fact, after these formal developments, we can ask: 
being parity \textit{compatible} with relativity, are the physical laws of nature
\textit{really} invariant under spatial inversion?

\noindent
The situation is different with respect to rotations, that represent a 
\textit{necessary} invariance for our understanding of nature.

\noindent
Initially, parity was considered an invariance of physics, but in the fifties
the situation changed.
In fact, some experiments on beta decay showed that 
\textit{weak interactions are not invariant} under spatial inversion.
On the other hand, \textit{gravitational, electromagnetic and strong (or nuclear)
interactions} are parity invariant.

\noindent
When deriving the Dirac equation, we shall require the fulfillment of parity invariance,
having in mind the study of electromagnetic and strong interactions.
In a following work we shall discuss the weakly-interacting neutrino equations, that are 
\textit{not invariant} under parity transformation. 

\vskip 0.5 truecm
\noindent
We conclude this section mentioning another discrete transformation, called 
\textit{time reversal}, that consists in changing the sign of time: $t'=-t$.
Classical laws of physics are invariant with respect to this change of the sense
of direction of time.
The action of time reversal on the space-time four vector is represented by the
matrix $T=-\Pi=-g$.

\vskip 0.5 truecm
\noindent
At microscopic level, time reversal invariance is \textit{exact}
in strong and electromagnetic 
processes, but not in weak interactions. However, this violation is of different kind
with respect to that of parity transformation. 

\noindent
We conclude pointing out that
in the formalism of field theories
the \textit{product} of the three transformations : 
C (\textit{Charge Conjugation}), P (\textit{Parity}) and T (\textit{Time Reversal}),
is an exact invariance, as confirmed by the available experimental data.
\section{Relativistic  Quantum Wave Equations}

\vskip 0.5 truecm
\noindent
In this Section we shall study the procedure to implement the principles of special 
relativity in the  formalism of quantum mechanics in order to introduce
the fundamental Dirac equation.

\noindent
Previously, in  Subsection 3.1 we shall analyze the general properties 
of the four-momentum operator
in quantum mechanics and discuss at pedagogical level the Klein-Gordon equation for
spinless particles.

\subsection{Generalities and Spin $0$ Equation}

\noindent
Let us firstly recall the Schr\"odinger equation for a free particle.
In the coordinate representation it has the form
$$i\hbar{{\partial \psi(t,{\bf r})}\over {\partial t}}=
-{\hbar^2\over {2 m}}\nabla^2\psi(t,{\bf r}) \eqno(3.1)$$
It can be obtained by means of the following eqs.(3.2a-c),
performing
the \textit{translation}, in terms of differential operators acting onto
the wave function $ \psi(t,{\bf r})$,
of the standard nonrelativistic expression
$$E ={{\bf p^2}\over {2m}}$$

\noindent
It  clearly shows  that Schr\"odinger equation (3.1)
is essentially nonrelativistic or, in other words,
\textit{ not compatible} with Lorentz transformations.
\vskip 0.5 truecm
\noindent
As discussed in refs.[5,6],
the fundamental relation that is used for the study of (relativistic) 
quantum mechanics associates
the four-momentum of a particle to a space-time differential operator in the
following form
$$p^{\mu}=i\hbar \partial^{\mu}\eqno(3.2a)$$
that, as explained in Subsection 2.1, means
$$p^0 c=E=i\hbar{\partial \over {\partial t}}\eqno(3.2b)$$
and
$${\bf p}=-i\hbar  \nabla=-i\hbar {{\partial \over\partial{\bf r}}}\eqno (3.2c)$$
The reader may be surprised that at relativistic level the \textit{same} relations 
hold as in nonrelativistic quantum mechanics. 
As a matter of fact, 
eqs.(3.2a-c) express  \textit{experimental } general properties of quantum waves, 
as given by the De Broglie hypothesis.

\noindent
Furthermore, the connection with relativity is possible because  
$i\hbar \partial^{\mu}$ is a contravariant four-vector operator.

\noindent
The easiest choice to write a relativistic wave equation consists in \textit{translating}
eq.(2.16) 
(\textit{instead of} the nonrelativistic  expression !) 
in terms of  the space-time differential operators  given 
by the previous equations. One has
$$-\hbar^2\partial_\mu\partial^\mu \psi(t,{\bf r})=(mc)^2\psi(t,{\bf r})\eqno(3.3a)$$
or, more explicitly, multiplying by $c^2$
$$-(\hbar c)^2\left({1\over c^2}{{\partial^2 \over\partial t^2}}- \nabla^2 \right)
\psi(t,{\bf r})
=m^2c^4\psi(t,{\bf r})\eqno(3.3b)$$
Exactly as done for the electromagnetic field equations in Subsection 2.3, 
recalling the invariance of $\partial_\mu \partial^\mu$,   one realizes
that previous equation is \textit{manifestly covariant}.

\vskip 0.5 truecm
\noindent
In order to make explicit calculations in atomic, nuclear and subnuclear physics,
it is necessary to remember   some numerical values 
(and the corresponding units !).
We start considering the following quantities that appear in eq.(3.3b):
$$\hbar c =197.327~MeV~fm$$
that is the Planck constant $\hbar$   multiplied by the speed of light $c$, expressed
as an energy multiplied by a length.
The energy is measured in $MeV$ 
$$1MeV=10^6 eV=1.6022 \times 10^{13} Joule$$ 
and the length in $fm$ 
\textit{(femtometers or Fermis)}
$$1 fm= 10^{-15}m=10^{-13}cm$$ 
Furthermore, the particle masses are conveniently expressed in terms of their
rest energies. We give a few relevant examples
$$m_e c^2=0.511~MeV$$
for the electron
$$ m_p c^2= 938.27 ~MeV$$
for the proton, and
$$ m_n c^2= 939.57 ~MeV$$
for the neutron.

\noindent
Also note that the operator $\partial^\mu$ is, dimensionally, a
$length^{-1}$, that in our units gives $fm^{-1}$.

\vskip 0.5 truecm
\noindent
Going back to the formal aspects of eq.(3.3a,b), usually called Klein-Gordon equation,
we note the two following aspects:

\vskip 0.5 truecm
\noindent
(i) Being based on the relativistic relation among energy, momentum and mass of 
eq.(2.16) with the De Broglie hypothesis of eqs.(3.2a-c), 
the \textit{manifestly covariant}  Klein-Gordon equation
has a \textit{general} validity, in the sense that
the wave fuctions of
\textit{all} the relativistic free particles  must
satisfy that equation.
As for the Dirac equation for spin $1/2$ particles, see eq.(3.41) 
and the following discussion.

\vskip 0.5 truecm
\noindent
(ii) In the Klein-Gordon equation it does not appear the particle spin. Or, equivalently,
the function $\psi(t,{\bf r})$ is a one-component or \textit{scalar} field function
that describes a spin $0$ particle, as it happens in nonrelativistic quantum mechanics
when spin is not included.

\noindent
The effects of rotations and Lorentz boosts only
consist in the \textit{argument re-expression} discussed 
in Subsection 2.2. As explained in textbooks of quantum mechanics,
see for example ref.[7],
the (infintesimal) rotations are performed by using the orbital angular
momentum operator as generator.

\vskip 0.5 truecm
\noindent
The Klein-Gordon equation admits plane wave solutions,
corresponding to  eigenstates of the four-momentum  $p^\mu=({E\over c},{\bf p})$ 
in the form
$$ \psi_{p}(t,{\bf r})=
N \exp\left[{i\over \hbar}(- E t +{\bf p r})\right]\eqno(3.4a)$$
$$ = N \exp\left( -{i\over \hbar}p_\mu x^\mu    \right)
   = N \exp\left( -{i\over \hbar} [p]^T g[x]   \right)
\eqno(3.4b)$$
where $N$ represents a normalization constant. The expression (3.4b) has 
been written using explicitly the Lorentz covariant notation.

\noindent
The most relevant point here is that the energy eigenvalue $E$ can assume 
both positive and negative values
(we shall see that it holds true also for Dirac equation !)
We have

$$E= p^0 c=\lambda \epsilon({\bf p})\eqno(3.5a)$$
where
$$\epsilon({\bf p})=[({\bf p} c)^2 + (mc^2)^2]^{1/2}\eqno(3.5b)$$
and the  \textit{energy sign} $\lambda=+/-1$ have been introduced.

\noindent
In quantum mechanics the $\lambda=-1$ solutions cannot be eliminated. They are strictly
necessary to have a \textit{complete} set of solutions of the wave equation.
They can be correctly interpreted by means of charge conjugation in the framework of
field theory, as done in most textbooks.
Historically, starting from the work by Dirac, negative energy solutions lead 
to the very important discovery of the \textit{antiparticles}, that have the same
mass (and spin) but \textit{opposite charge} with respect to the corresponding particles.

\noindent
We shall not analyze this problem here but postpone it to a subsequent work.

\vskip 0.5 truecm
\noindent
As for the positive energy solutions, one can immediately check that in the
nonrelativistic regime ($|{\bf p}| c << mc^2$) the Schr\"odinger limit is obtained.

\vskip 0.5 truecm
\noindent
As an illustrative exercise, it may be useful to perform a Lorentz boost in eq.(3.4b).
Given that we are considering a scalar field, we have to make only the
\textit{argument re-expression}.

\noindent
In this concern
recall that, for positive energy, the wave function of eq.(3.4a,b) 
represents a particle state
such that an observer in $\cal S $ measures  the particle
four-momentum $p^\mu$.

\noindent
In the reference frame $\cal S'$, for the space-time position one must use
$$ [x]= L^{-1}[x']$$
(both the velocity and the hyperbolic parametrizations can be adopted 
and, for simplicity, no argument has been written in  $L^{-1}$)
and replace it in eq.(3.4b). In the argument of plane wave exponential one has

$$[p]^T g[x]  = [p]^T gL^{-1}[x']=[p]^T L g [x']=[p']^T g [x']$$
where we have used $ gL^{-1}=Lg $ from eq.(2.10) and also
$[p]^T L=[p']^T $.
In the previous result we recognize the invariance equation that, 
in standard notation, reads
$$p_\mu x^\mu=p'_\mu x'^\mu$$
\textit{Physically},
it  means that an observer in $\cal S'$  measures the particle transformed four-momentum
$p'^\mu $.

\vskip 0.5 truecm
\noindent
The Klein-Gordon equation admits a \textit{conserved current}.
We shall consider the form, related to a \textit{transition process}, that is used 
in perturbation theory to calculate the corresponding probability amplitude.

\noindent
To derive the conserved current one has to make the following  three steps.

\vskip 0.5 truecm
\noindent
(i) Take eq.(3.3b) with a plane wave solution $\psi_{p_I}(t,{\bf r})$
 for an initial state  of four-momentum $p_I$.

\vskip 0.5 truecm
\noindent
(ii) Take eq.(3.3b) with a plane wave solution $\psi_{p_F}(t,{\bf r})$
 for a final state  of four-momentum $p_F$ and make the complex conjugate.

\vskip 0.5 truecm
\noindent
(iii) Multiply the equation of step (i) by
the complex conjugate
$\psi^*_{p_F}(t,{\bf r})$ and 
the equation of
step (ii) by  $\psi_{p_I}(t,{\bf r})$. Then subtract these 
two equations, obtaining
$$[\partial^\mu \partial_\mu \psi^*_{p_F}(t,{\bf r})]\psi_{p_I}(t,{\bf r})-
\psi^*_{p_F}(t,{\bf r})\partial_\mu\partial^\mu\psi_{p_I}(t,{\bf r}) =0
\eqno(3.6)$$
Note that the mass term has disappeared.
The previous equation can be equivalently written as a 
\textit{conservation equation} in the form
$$\partial_\mu J^\mu_{F I}(t,{\bf r})=0 \eqno(3.7)$$
where the \textit{conserved current} is defined as
(multiplying by the conventional factor $i\hbar$)
$$J^\mu_{F I}(t,{\bf r})=i\hbar[
\psi^*_{p_F}(t,{\bf r})\partial^\mu \psi_{p_I}(t,{\bf r})-
(\partial^\mu\psi^*_{p_F}(t,{\bf r}))\psi_{p_I}(t,{\bf r})]\eqno(3.8a)$$
$$=(p_I^\mu+ p_F^\mu)N_I N_F 
\exp\left( {i\over \hbar}q_\mu x^\mu    \right)\eqno(3.8b)$$
In the last equation the four-momentum transfer $q^\mu=p_F^\mu-p_I^\mu $ 
of the transition process has been introduced.

\noindent
The conserved current   $J^\mu_{F I}(t,{\bf r})$ is \textit{manifestly} a four-vector.
\vskip 0.5 truecm
\noindent
The latter eq.(3.8b), that is obtained by explicit use of the
wave functions, is very interesting. The first term  $(p_I^\mu+ p_F^\mu)$
represents the so-called  four-vector \textit{vertex factor}. 

\noindent 
Applying to eq.(3.8b) the derivative operator 
$\partial_\mu$ one verifies that current conservation relies
on the following \textit{kinematic} property of the vertex factor
$$q_\mu (p_I^\mu+ p_F^\mu)= 
{p_F}^\mu {p_F}_ \mu -   {p_I}^\mu {p_I}_\mu=  0\eqno(3.9)$$
that is \textit{automatically} satisfied because
the mass of the particle  remains \textit{the same} in the initial and final state.

\vskip 0.5 truecm
\noindent
As for the general properties of the current  given in eqs.(3.8a,b) we find that
in the \textit{static case}, i.e. $p_F=p_I$, 
the time component  $J^0_{I I}$ is \textit{negative} if negative energy states
$(\lambda =-1)$ are  considered. 
It means that one cannot 
attach to $J^0_{I I}$ the meaning of \textit{probability density} as it was done with
the Schr\"odinger equation. 
For this reason we do not discuss in more detail the plane wave normalization 
constant $N$.

\noindent
Again, a complete interpretation of the Klein-Gordon equation and of its current
is obtained in the context of field theory. 
\subsection{Spin $1/2$ Dirac Equation}

\noindent
In nonrelativistic quantum-mechanics  a spin $1/2$ particle is described by a
\textit{two-component} spinor $\phi$. 
The spinor rotation is performed by mixing its components. At first order in the
rotation angle ${\bf \theta^\alpha}$, one has
$$\phi'\simeq( 1-{i\over 2} \theta^\alpha \sigma^\alpha)\phi \eqno(3.10)$$
where the three Pauli matrices $\sigma^\alpha$ have been introduced. 
Their properties are studied in the Appendix.

\noindent
What is important to note here is that  the matrix operators
 $S_{[2]}^\alpha=-{i\over 2}  \sigma^\alpha $    play the same r\^ole 
in realizing the rotations as the matrices $S^\alpha$
defined in   eq.(2.49). 
For this reason, their
commutation rules  are the same as those given in eq.(2.50b). 
Also, the \textit{spin} or \textit{intrinsic angular momentum} operator is defined 
[7]  multiplying by $\hbar/2$ the Pauli matrices $\sigma^\alpha$.

\noindent
Formally, we have introduced the two-dimensional representation of the rotation group
( the three-dimensional representation corresponds to spin 1, etc.).

\noindent
Finally, the spatial \textit{argument} of the spinor $\phi$ 
(not written expicitly in eq.(3.10)) is rotated with the same rules previously discussed
for the arguments of the field functions, 
that is one has to perform the \textit{argument re-expression}.

\noindent
In quantum mechanics, the generator of these
rotations is the orbital angular momentum operator $ {\bf l=r\times p}~$, so that the
total angular momentum is given by the three generators of the total rotation 
(on the spinor and on the argument), in the form
$$j^\alpha= l^\alpha+{\hbar\over 2}\sigma^\alpha$$

\vskip 0.5 truecm
\noindent
We can now try to introduce relativity. 
We shall follow a strategy similar to that of refs.[5,8], but avoiding many
unessential (at this level) mathematical details.

\noindent
First, we note that for a particle \textit{at rest}, the relativistic theory must
coincide with the previous nonrelativistic treatment.

\noindent
Second, we make the following question: can we find a set of three $2\times2$ 
boost matrices
(acting on the two-component spinors) that satisfy, 
with the $S_{[2]}^{\alpha}=-{i\over 2}  \sigma^\alpha $ replacing the $S^\alpha$, 
the same commutation rules as the $K^\alpha$ in eqs.(2.50a-c)? 

\noindent
The answer is \textit{yes}. A simple \textit{inspection} of eqs.(2.50a-c) and use
of the standard property of the Pauli matrices given in eq.(A.2)  show that the matrices
$K_{[2]}^\alpha = {\tau\over 2}\sigma^\alpha$ 
satisfy those commutation rules.

\noindent
Eq.(2.50a) requires $\tau^2=1$, while eq.(2.50c) does not give any new constraint on
the parameter $\tau$, that, in consequence can be chosen equivalently as 

\noindent
$\tau=+/-1$.
 
\vskip 0.5 truecm  
\noindent
However, a serious problem arises when trying to introduce the parity transformation 
matrix. It must satisfy,  both the anticommutation rule with 
the boost generators as in eq.(2.51a) and the commutation rule with the rotation 
generators as in eq.(2.52a). In our $2\times 2$ case, boost generators
and rotation generators are proportional
to the Pauli matrices, so there is no matrix that satisfies
\textit{at the same time}
the two rules [8].

\noindent
In consequence, we can construct a two-dimensional theory for spin $1/2$ particles 
that is invariant under Lorentz transformations but not under parity transformations.

\noindent
On the other hand, the first objective that we want to reach is the study of 
the electromagnetic interactions of the electrons in atomic physics and in scattering 
processes. 
To this aim we need an equation that is invariant under spatial inversion. 

\noindent
A parity noninvariant equation for spin $1/2$ particles,
based on the transformation properties outlined above,
will be used for the study of
the \textit{neutrinos} that are created, destroyed and in general interact only by means
of \textit{weak interactions} that are not invariant under spatial inversions.

\vskip 0.5 truecm
\noindent
In order to construct a set of matrices for spin $1/2$ particles  satisfying both 
Lorentz and parity commutation rules, we make the two following steps:

\noindent
(i) we consider matrices with \textit{larger} dimension;

\noindent
(ii) we exploit the sign ambiguity of $\tau$ in the boost generator.

\noindent
More precisely, it is sufficient to introduce the  following $4\times 4$ 
\textit{block} matrices

$$K_{[D]}^\delta={1\over 2}\left[ \matrix{ \sigma^\delta ~~~~0 \cr
                                          ~~0  ~-\sigma^\delta}\right]
=  {1\over 2}\alpha^\delta \eqno(3.11)$$
where we have taken $\tau=+1$ and   $\tau=-1$  in the upper and lower diagonal block,
respectively. 

\noindent
Important note: the previous equation represents the definition of the three
matrices $\alpha^\delta$. 
We use the greek letter $\delta$ (instead of $\alpha$) as spatial 
index to avoid confusion between the indices and the matrices. 

\vskip 0.5 truecm
\noindent
With no difficulty, for the spinor rotations we introduce
$$\Sigma^\delta=\left[ \matrix{ \sigma^\delta ~~0 \cr
                                          ~0  ~~~~\sigma^\delta}\right]\eqno(3.12a)$$
so that
$$S_{[D]}^\delta=-{i\over 2}\Sigma^\delta\eqno(3.12b)$$

\noindent
Note that, taking into account the discussion for the transformation of  
the two-dimensional spinors with $S_{[2]}^\delta$ and $K_{[2]}^\delta$, 
the commutation rules of eqs.(2.50a-c) for Lorentz transformations and rotations
are automatically satified by the block diagonal matrices
$K_{[D]}^\delta$,  $S_{[D]}^\delta$ introduced above.

\vskip 0,5 truecm
\noindent
For the spatial inversion, we find the $4\times 4$ block matrix
$$\Pi_{[D]}=\left[ \matrix{ 0 ~~  1 \cr
                            1 ~~ 0}\right]=\beta\eqno(3.13)$$
with the property
$$\Pi_{[D]}=   \Pi_{[D]}^{\dagger}=   \Pi_{[D]}^{-1}\eqno(3.14)    $$
It satisfies the anticommutation with the boost generators of eq.(2.51a,b),
that means
$$\{\Pi_{[D]},K_{[D]}^\delta \}=  \{\beta,\alpha^\delta \}=   0\eqno(3.15)$$
\noindent
The specific form of $\Pi_{[D]}$ straightforwardly  satisfies also
the rules (2.52a,b).

\noindent
In technical words, we have obtained a \textit{representation} of the Lorentz group, 
including parity, for spin $1/2$ particles.

\vskip 0.5 truecm
\noindent
Introducing explicitly the four component Dirac spinor $u$,
its boost transformation is written in the form
$$u'=B_{[D]}(\omega)u\eqno(3.16)$$

\noindent
The (infinitesimal) form of $B_{[D]}(\omega)$  at first order in 
$\omega$ is
$$B_{[D]}(\omega)\simeq 
1 -{1\over 2}~\omega~( \alpha {\bf \hat v})\eqno(3.17a) $$
where, as usual, ${\bf \hat v}$ represents the unity vector of the boost velocity. 
(For simplicity, we do not write it explicitly in  $B_{[D]}(\omega)$.)

\noindent
The \textit{finite} transformation is obtained in the same way as in 
eqs.(2.36a,b) and (2.40) but using the properties of the Pauli matrices,
as it is shown in detail in eqs.(A.17),(A.18) and in the following discussion 
in  the Appendix. One has
$$B_{[D]}(\omega)=\exp[-{\omega\over 2}( \alpha {\bf \hat v})]=
 ch({\omega\over 2})
-( \alpha {\bf \hat v}) sh({\omega\over 2}) \eqno(3.17b)$$
On the other hand, the spinor rotations are obtained by replacing the $\sigma^\delta$
with the  $4\times 4$ matrices $\Sigma^\delta$ in eq.(3.10).

\noindent
Furthermore, when changing the reference frame,
one has always to perform the \textit{argument re-expression} 
in the Dirac spinors $u$.

\noindent
We note that, while the rotations are represented by a \textit{unitary operator},
the Lorentz boost are \textit{not}. More precisely, $B_{[D]}(\omega)$
is a \textit{antiunitary operator}, that is
 $$B_{[D]}^{\dagger}(\omega)=  B_{[D]}(\omega)\eqno(3.18)$$
A unitary, but infinite dimensional (or nonlocal) representation of the boost
for spin $1/2$ particles can be obtained. 
This problem will be studied in a different work.

\vskip 0.5 truecm
\noindent
The next task is to construct \textit{matrix elements} 
(in the sense of vector algebra and not of quantum mechanics,
because no spatial integration is performed) of the form
$~{u_b}^{\dagger} M u_a ~ $, that, when boosting $u_a$ and $u_b$, transform as
Lorentz scalar and Lorentz four-vectors. The case of pseudoscalars and axial-vectors
will be studied in Subsection 3.4. 

\noindent
We shall keep using the word \textit{matrix elements} throughout
this work, but in most textbooks 
they are commonly denoted as
\textit{ Dirac covariant bilinear} quantities.

\vskip 0.5 truecm
\noindent
Given a generic $4\times 4$ matrix $M$, by means of eq.(3.17a) the transformation
of the matrix element up to first order in $\omega$, is
$${u_b'}^{\dagger} M u_a' \simeq  {u_b}^{\dagger} M u_a 
-{1\over 2}~\omega~{\hat v}^\delta  
{u_b}^{\dagger} \{\alpha^\delta, M \} u_a \eqno(3.19)$$
The Lorentz scalar matriz element is easily determined by means of a matrix $M_s$
that anticommutes with the $\alpha^\delta$ so that the  second term  in the 
\textit{r.h.s.} of eq.(3.19) is vanishing. 
Simply recalling eqs.(3.15) and (3.13) one has
$$M_s=\beta=\left[ \matrix{ 0 ~~  1 \cr
                            1 ~~ 0}\right]\eqno(3.20)$$
where we are using the definition of the $\beta$ Dirac matrix given in eq.(3.13).

\vskip 0.5 truecm
\noindent
As for the four-vector matrix element, one needs four matrices $M_v^\mu$. 
To find their form in a simple way, let us consider a boost along the
$x^1$-axis, that in eq.(3.19) means ${\bf \hat v}=(1,0,0)$.
By means of eq.(3.19),
to recover the four-vector Lorentz transformation (see eqs.(2.1) and (2.34)),
one needs
$${1\over 2}\{\alpha^1,M_v^0 \}=M_v^1 \eqno(3.21)$$
for the transformation of $M_v^0$, and
$${1\over 2}\{\alpha^1,M_v^1 \}=M_v^0 \eqno(3.22)$$
for the transformation of $M_v^1$.

\noindent
The solution is easily found calculating the anticommutators of 
the Dirac matrices $\alpha^\delta$  
by means of the anticommutators of the Pauli matrices of eq.(A.3). One has 
$$M_v^0= \alpha^0= 1,~~~M_v^1=\alpha^1 \eqno(3.23a)$$
and the solution for all the components is
$$M_v^\mu=\alpha^\mu=(1,\alpha^1,\alpha^2,\alpha^3) \eqno(3.23b)$$
Pay attention: $\alpha^0=1$ is not introduced in most textbooks.

\vskip 0.5 truecm
\noindent
We can resume the previous equations, also for finite Lorentz boosts, as
$$B_{[D]}(\omega)\beta B_{[D]}(\omega)= \beta \eqno(3.24a)$$
or, equivalently
$$B_{[D]}(\omega)\beta = \beta  B_{[D]}^{-1}(\omega)\eqno(3.24b)$$
for the scalar matrix elements, and
$$B_{[D]}(\omega)\alpha^\mu B_{[D]}(\omega)= 
L^\mu_{~\nu}(\omega) \alpha^\nu   \eqno(3.25)$$
for the four-vector ones.

\vskip 0.5 truecm
\noindent
The previous developments, recalling the expression of the four-momentum operator
given in eqs.(3.2a-c), allow to write a \textit{linear} covariant wave equation
in the form
$$i\hbar c ~ \partial_\mu \alpha^\mu \psi(x)=mc^2 ~ \beta \psi(x) \eqno(3.26)$$
that is the Dirac equation, where $m$ is the particle mass and 
$\psi(x)=\psi(t,{\bf r})$ is a four component Dirac spinor representing the
particle wave function.

\noindent
Intuitively, the covariance of the Dirac equation can be proven multiplying the
previous equation from the left by a generic hermitic conjugate  Dirac spinor.
In the \textit{l.h.s.} one has a Lorentz scalar given by the product of the
(contravariant) four-vector martix element of $\alpha^\mu$ with the
(covariant) operator $ i\hbar c ~ \partial_\mu ~$.
In the \textit{r.h.s.} one has the Lorentz scalar directly given by the matrix
element of $\beta$.

\noindent
More formally, we can prove the covariance of the Dirac equation in the following way.
We write the same equation in $\cal S'$ and show that is equivalent to the 
(original) equation in $\cal S$. We have

$$i\hbar c ~ \partial~'_\mu \alpha^\mu \psi'(x')=mc^2 ~ \beta \psi'(x') \eqno(3.27)$$
The spinor in $\cal S'$ is related to the spinor in $\cal S $ by means of eq.(3.16):
$$\psi'(x') =B_{[D]}(\omega)\psi(x'(x)) \eqno(3.28)$$
We replace the last expression in eq.(3.27) and multiply from the left that equation by 
$B_{[D]}(\omega)$. In the \textit{r.h.s.}, by means of eq.(3.24a) one directly obtains
$\beta \psi$. In the \textit{l.h.s.},  one has to consider eq.(3.25), transforming 
the equation in the form
$$ i\hbar c ~  \partial~'_\mu L^\mu_{~\nu}(\omega) \alpha^\nu \psi(x'(x))=
mc^2 ~ \beta \psi(x'(x))$$
We can use the more synthetic vector algebra notation, writing 
$$\partial~'_\mu L^\mu_{~\nu} \alpha^\nu =[\partial~']^T g L [\alpha]
=[\partial]^T g  [\alpha]=\partial_\mu \alpha^\mu \eqno(3.29)$$
where in the second equality we have used $g L =L^{-1}g $.

\noindent
In this way we have shown
the equivalence of eq.(3.27), written in $\cal S'$,
with the original equation (3.26), written in $\cal S$.

\subsection{The Gamma Dirac Matrices and the Standard Representation}

\noindent
The physical content of the Dirac equation is completely contained in eq.(3.26) 
and in the related transformation properties.
However, to work in a more direct way with Dirac equation and its applications, 
some more developments are necessary.

\noindent
First, we introduce the \textit{Dirac adjoint} spinor that is preferably used 
(instead of the hermitic conjugate) to calculate matrix elements. It is defined as
$$\bar u=u^\dagger \beta \eqno(3.30)$$
Its transformation law is straightforwardly obtained 
in the form
$$\bar u'=u'^\dagger\beta=u^\dagger B_{[D]}(\omega)\beta =
\bar u B_{[D]}^{-1}(\omega) \eqno(3.31) $$
where eq.(3.24b) has been used.
As it must be for a representation of the Lorentz boost, $B_{[D]}^{-1}(\omega) $
is obtained inverting the direction of the boost velocity
$$B_{[D]}^{-1}(\omega)= ch({\omega\over 2})
+(\alpha {\bf \hat v}) sh({\omega\over 2}) \simeq
1 +{1\over 2}~\omega~   (\alpha {\bf \hat v})  \eqno(3.32) $$
As an exercise, the reader can check that $B_{[D]}(\omega)B_{[D]}^{-1}(\omega)=1$
by using the properties of the $\alpha^\delta$ matrices.

\noindent
Note that in the previous results there is no new physical content. We can represent 
the Lorentz scalar (\textit{invariant}) as
$$u^\dagger_b \beta u_a=\bar u_b u_a \eqno(3.33)$$
In fact, we have learned in eq.(3.31) that $\bar u$ transforms with $B_{[D]}^{-1}(\omega)$.

\vskip 0.5 truecm
\noindent
We introduce the Dirac  matrices $\gamma^\mu$  defined as
$$\gamma^\mu=\beta \alpha^\mu \eqno(3.34a)$$
Recalling that $\beta^2=1$, one  has
$$\alpha^\mu=\beta \gamma^\mu  \eqno(3.34b)$$
The four-vector matrix element  can be written as
$$u_b^\dagger \alpha^\mu u_a=\bar u_b\gamma^\mu u_a \eqno(3.35)$$
and the Dirac equation  (3.26) takes the usual form

$$i\hbar c ~ \partial_\mu \gamma^\mu \psi(x)=mc^2 ~  \psi(x) \eqno(3.36)$$
For clarity we give the explicit  expression of the $\gamma^\mu$ :
$$\gamma^0=\beta=\left[ \matrix{ 0 ~~  1 \cr
                                  1 ~~ 0}\right],~~~
 \gamma^\delta=\left[ \matrix{ ~~0  ~-\sigma^\delta \cr
                               \sigma^\delta ~~~0}
\right]\eqno(3.37)$$
As it will be discussed in the following, this is the so-called
\textit{spinorial representation} of the Dirac matrices.

\noindent
Starting from the anticommuation rules of the $\alpha^\mu$ one finds the following
\textit{fundamental !} anticommutation rules of the $\gamma^\mu$
$$\{\gamma^\mu,\gamma^\nu \}=2g^{\mu \nu} \eqno(3.38)$$
Furthermore, one easily verifies that $\gamma^0 $ is hermitic while the
$\gamma^\delta$ are anti-hermitic:

$${\gamma^\mu}^\dagger=g^{\mu \mu}\gamma^\mu=
\gamma^0 \gamma^\mu \gamma^0 
\eqno(3.39)$$  
Note that in $g^{\mu \mu}$  the index $\mu$ not summed; the last equality is obtained
by standard use of eq.(3.38). Furthermore, the previous equation also holds in
the \textit{standard representation} of the Dirac matrices that will be introduced 
in the following.

\vskip 0.5 truecm
\noindent
We can now easily examine the \textit{usual} procedure that is adopted to introduce the
Dirac equation.
Consider, for example, refs.[6,9].
The differential wave equation for a spin $1/2$ particle is assumed to be \textit{linear} 
with respect to the
four-momentum operator introduced in  eq.(3.2a,c) and to the particle mass.
According to this hypothesis, the equation is written as

$$i\hbar c ~ \partial_\mu \Gamma^\mu \psi(x)=mc^2 ~  \psi(x) \eqno(3.40)$$
where the $\Gamma^\mu$ are four adimensional matrices to be determined. 

\noindent
Then, one multiplies by $i\hbar c ~ \partial_\mu \Gamma^\mu $ and, by using the same
eq.(3.40), obtains in the \textit{l.h.s.} another  factor $mc^2$. 
The equation takes the form
$$-(\hbar c)^2 \partial_\nu \Gamma^\nu  \partial_\mu \Gamma^\mu \psi(x)=
(mc^2)^2 ~  \psi(x) \eqno(3.41)$$
As we said in Subsection 3.1,  the wave function of \textit{any} relativistic particle
must satisfy the Klein-Gordon equation (3.3a,b). This property must be verified also
in our case. To this aim, we make the following algebraic manipulation
$$\partial_\nu \Gamma^\nu  \partial_\mu \Gamma^\nu=
{1\over 2}~ \partial_\mu \partial_\nu 
(\Gamma^\nu \Gamma^\nu + \Gamma^\mu \Gamma^\nu ) $$
It shows that the $\Gamma^\mu$ must satisfy the anticommutation rules of eq.(3.38).
The lowest dimension for which it is possible is 4 and we can identify the 
$\Gamma^\mu$  with the $\gamma^\mu$ of eq.(3.37) that have been derived by means
of relativistic transformation properties.

\noindent
In any case, (we repeat)  the previous development is useful to show
 that the solutions of the Dirac equation  are
also solutions of the Klein-Gordon one. 
We can expect that also Dirac equation admits negative energy solutions.

\vskip 0.5 truecm
\noindent
We now face a different problem. In Subsection 3.2 we have seen that the relevant 
point for the covariance of the Dirac equation is represented by the anticommutation 
rules of the $\alpha^\delta$ and $\beta$ matrices. The same is true for the $\gamma^\mu$.
In other words, their specific form is not important, provided that the anticommutation 
rules are fulfilled. We now look for another representation, different from eq.(3.37), and
more useful for practical calculations.
We construct this new representation starting from a specific solution of Dirac equation
(3.26) or (3.36).

\noindent
Let us consider a particle at rest, that is, in a three-momentum eigenstate with 
${\bf p} =0$. The spatial components ${\partial \over {\partial {\bf r}}}$ 
of the derivative operator, when applied to the corresponding wave function, give zero.
The Dirac equation reduces to
$$i\hbar {{\partial \psi(x)}\over {\partial t}}=
mc^2 \left[ \matrix{ 0 ~~  1 \cr
                            1 ~~ 0}\right]\psi(x)
\eqno(3.42a)$$
We can split the Dirac spinor into two two-component spinors
$$\psi=\left[ \matrix{ \eta  \cr
                       \xi }\right]$$     
So that eq.(3.42a) is written as a system of coupled equations:
$$i\hbar {{\partial \eta}\over {\partial t}}=mc^2 \xi $$
$$i\hbar {{\partial \xi }\over {\partial t}}=mc^2 \eta \eqno(3.42b)$$
We can sum and subtract these two equations  introducing the new 
two-component spinors
$$\varphi={1\over \sqrt{2}}(\xi+\eta)$$
$$\chi   ={1\over \sqrt{2}}(\xi-\eta)\eqno(3.43)$$
(the factor ${1\over \sqrt{2}}$ guarantees that normalization of the new Dirac 
spinor is not changed).
One finds
$$i\hbar {{\partial \varphi }\over {\partial t}}=mc^2 \varphi$$
$$i\hbar {{\partial \chi }\over {\partial t}}=-mc^2 \chi \eqno(3.44)$$
These equations are equivalent to eq.(3.42b) but they are \textit{decoupled}.
Technically, we have diagonalized the \textit{r.h.s.} rest frame Hamiltonian 
of eq.(3.42a).

\noindent
The solutions are easily found:
$$\psi_+=\left[ \matrix{ \varphi  \cr
                       0 }\right]$$
with \textit{positive energy} $E=+mc^2$,
$$\psi_-=\left[ \matrix{ 0  \cr
                       \chi }\right]$$
with \textit{negative energy} $E=-mc^2$. 
The presence of two energy values  represents a general property
of relativistic wave equations.

\noindent
The advantage of the solutions $\psi_{+/-}$ of eq.(3.44) is that 
\textit{only one} two-component spinor is nonvanishing while the other is zero.
In the positive energy case, the nonvanishing spinor can be identified with the 
nonrelativistic one. Furthermore, when considering a positive enery particle 
with small (nonrelativistic) velocity, we can expect  the lower components
of $\psi_+$ to be (not zero but) small with respect to the upper ones.

\noindent
For these reasons we apply that transformation to a generic Dirac spinor, not only
in the case ${\bf p=0}$.

\noindent
More formally, we perform the transformation of eq.(3.43) by introducing 
the following matrix
$$U= {1\over \sqrt{2}} \left[ \matrix{~ 1 ~~~~  1 \cr
                            ~1 -1}\right]\eqno(3.45)$$
that satisfies 
$$U^\dagger=U^{-1}=U$$
We multiply  from the left the Dirac equation (3.36) by U and insert
$UU=1$ between the $\gamma^\mu$  and $\psi$ .
In this way we transform the Dirac wave function
and, at the same time, the Dirac matrices obtaining
$$\gamma_{st}^{\mu}=U\gamma^{\mu}U\eqno(3.46)$$
where the  $\gamma_{st}^{\mu}$ are the Dirac matrices in the 
\textit{standard representation}, while the $\gamma^\mu$ of eq.(3.37) have been
 given in the so-called \textit{spinorial representation}.

\noindent
In most physical problem (specially if a connection with nonrelativistic physics
is wanted) the \textit{standard representation} is adopted.
Generally  the index \textit{``st"} is not explicitly written. 
In the following of the present work  we shall also adopt this convention.

\vskip 0.5 truecm
\noindent
Note that, due to the property of $U$  given above, if two matrices 
in the spinorial representation   satisfy an
(anti)commutation rule, the corresponding matrices
in the standard representation also satify the same rule.
  
\vskip 0.5 truecm
\noindent
In particular, this property holds for the anticommutation rule of eq.(3.38) 
of the $\gamma^{\mu}$.
In the  standard representation they have the form

$$\gamma^0=\beta=\left[ \matrix{ 1 ~~~~  0 \cr
                                  0 ~ -1}\right],~~~
 \gamma^\delta=\left[ \matrix{ ~~~0 ~~~~\sigma^\delta \cr
                               -\sigma^\delta ~~~0}
\right]\eqno(3.47)$$
The hermitic conjugate satify the same eq.(3.39).
As for the $\alpha ^\mu$, by using eq.(3.34b), one has
$$\alpha_{st}^\mu=U \alpha^\mu U= U\beta U U\gamma^\mu U=
\gamma_{st}^0\gamma_{st}^\mu \eqno(3.48)$$
Explicitly, \textit{ without writing the index ``st''}, they are
$$\alpha^0   =\left[ \matrix{ 1 ~~~~  0 \cr
                               0 ~~~~1}\right],~~~   
\alpha^\delta=\left[ \matrix{ ~~~0 ~~~~\sigma^\delta \cr
                              ~ \sigma^\delta ~~~0}
\right]\eqno(3.49)$$
Note that the spin $\Sigma^\delta$ matrices of eq.(3.12a) keep
the same form in the spinorial and standard representation. 

\noindent
In consequence, one can define $ K_{[D]}={1\over 2}\alpha^\delta$ and
$S_{[D]}^\delta=-{i\over 2}\Sigma^\delta$ by using the standard representation for the
$\alpha^\delta$ (and the $\Sigma^\delta$): the boost and rotation generators commutation
rules are equivalently fulfilled.
Furthermore, the expression of the boost operator is the same as in eq.(3.17a,b),
with the $\alpha^\delta$ written in the standard representation. 

\subsection{Parity Transformations and the Matrix $\gamma^5$}
\noindent
There is a fifth matrix that anticommutes with the other $\gamma^\mu$. It is
$\gamma^5$:

$$\{ \gamma^\mu,\gamma^5 \}= 0 \eqno(3.50)$$
In the spinorial and standard representations, one has, respectively
$$\gamma_{sp}^5  =\left[ \matrix{ -1 ~~~~  0 \cr
                               ~~0 ~~~~1}\right],~~~~~
  \gamma_{st}^5  =\left[ \matrix{~~ 0 ~  -1 \cr
                               -1 ~~~~0}\right] \eqno(3.51)$$
Note that ${\gamma^5}^\dagger= \gamma^5$ and $(\gamma^5)^2=1$.

\noindent
Furthermore, we use the definition of ref.[5], but, 
 as done in many texbooks, $\gamma^5$ can be defined multiplying
eq.(3.51) by $-1$. All its properties remain unchanged.
Pay attention to which definition is used !

\vskip 0.5 truecm
\noindent
To understand the physical meaning of the matrix elements of $\gamma^5$, it is
useful to go back to Dirac spinor parity transformation.
As shown in eq.(3.13), this transformation is $u'=\beta u$ being $\beta=\gamma^0$.
Let us consider the parity transformation for Lorentz scalar  and four-vector
matrix elements.
Standard use of the $\gamma^\mu$ anticommutation rule (3.38) gives
$$\bar {u_b}' {u_a}' = \bar u_b u_a \eqno(3.52a)$$
and
$$\bar {u_b}' \gamma^0 {u_a}' = \bar u_b \gamma^0 u_a \eqno(3.52b)$$
$$\bar {u_b}' \gamma^\delta {u_a}' = -\bar u_b \gamma^\delta u_a \eqno(3.52c)$$
These results have an easy physical interpretation:
a Lorentz scalar and a time component of a four-vector (for example a charge density)
do not change sign under spatial inversion, while the spatial components of
a four-vector (for example a current density) do change sign.

\vskip 0.5 truecm
\noindent
Let us now consider the following matrix element
$$\bar u_b  \gamma^5 u_a $$
The Lorentz boost are studied by means of eq.(3.19) taking $M_{ps}=\gamma^0\gamma^5$.
Standard use of eqs.(3.50)  (3.38) and (3.34a,b) show that
$$\{ \alpha^\delta, \gamma^0\gamma^5 \}=0\eqno(3.53)$$
so that we can conclude that our matrix element is \textit{invariant} under 
Lorentz transformations. The same can be shown for rotations using 
the generator of eq.(3.12a,b).

\noindent
But, what happens with spatial inversion  ? We have
$$\bar {u_b}' \gamma^5 {u_a}' = 
\bar {u_b}'\gamma^0 \gamma^5 \gamma^0 {u_a}' = - \bar u_b u_a \eqno(3.54)$$
It means that our matrix element changes sign under parity transformation.
It is a \textit{pseudo-scalar} quantity.

\noindent
In terms of elementary quantities, a pseudo-scalar is given by the product of 
an axial vector (see the discussion of subsect 2.5) with a standard vector,
for example the spin with the three-momentum: ${\bf s p}$. (It is not possible to use 
the orbital angular momentum instead of spin because one has  ${\bf l p}= 0$,
identically).

\vskip 0.5 truecm
\noindent
We now consider the following matrix element
$$\bar u_b  \gamma^5 \gamma^\mu u_a $$
Standard handling (that is left as an exercise)
 with the $\gamma^\mu$ and $\gamma^5$ shows that, under Lorentz boosts
and rotations, it transforms as a four-vector, but, under spatial inversion, one has
$$\bar {u_b}'  \gamma^5 \gamma^0 {u_a}' =
 \bar u_b  \gamma^0 \gamma^5 \gamma^0 \gamma^0 u_a =
-\bar u_b  \gamma^5 \gamma^0 u_a\eqno(3.55a)$$
and
$$\bar {u_b}'  \gamma^5 \gamma^\delta {u_a}' =
 \bar u_b  \gamma^0 \gamma^5\gamma^\delta  \gamma^0 u_a =
+\bar u_b  \gamma^5 \gamma^\delta u_a\eqno(3.55b)$$
We have an \textit{axial four-vector}. Its time component  changes sign,
while the space components do not.
\subsection{Plane Wave Solutions and the Conserved  Dirac Current }
\noindent
In this last Subsection we shall find the plane wave solutions of the Dirac equation
for a noninteracting particle, and, as in the case of the Klein-Gordon equation,
we shall determine the conserved current.

\vskip 0.5 truecm
\noindent   
At this point the equations become very large and it is necessary to find a strategy
to simplify the calculations and avoid to lose the physical meaning of the developments.
For this reason, most textbooks adopt the system of units in which
$$\hbar=c=1$$
In any part of the calculations one can go back to the standard units recalling the
following dimensional equalities
$$[\hbar]=[E]~[T],~~~~~[c]=[L]~[T]^{-1} $$
and use the numerical values given in Subsection 3.1.

\noindent
In this way, Dirac equation (3.36) is written in the form
$$[i \partial_\mu \gamma^\mu -m] ~  \psi(x)=0 \eqno(3.56)$$
Let us make the hypothesis that the wave function $\psi(x)$ can be factorized in 
plane wave exponential, identical to that of the Klein-Gordon equation given 
in eqs.(3.4a,b), and a Dirac spinor \textit{not depending} on 
the four-vector $x$.
Also using eq.(3.5a) for positive and negative energy,
being $\lambda$ the energy sign, we can write
$$\psi_{\lambda {\bf p} \sigma}(x)= u(\lambda,{\bf p},\sigma)
\exp\left[ i(-\lambda \epsilon({\bf p})t +{\bf pr})\right] \eqno(3.57)$$
The spin label $\sigma$ of the Dirac spinor (not to be confused with the Pauli matrices)
will be  discussed in the following.

\noindent
Applying the space-time derivative operator to the previous equation one has
$$i\partial_{\mu}\psi_{\lambda {\bf p} \sigma}(x)=
(\lambda \epsilon({\bf p}),-{\bf p})u(\lambda,{\bf p},\sigma)
\exp\left[ i(-\lambda \epsilon({\bf p})t +{\bf pr})\right] \eqno(3.58)$$
where the the minus sign in $-{\bf p}$ is due to the use of covariant components
of the operator $i\partial_\mu$.

\noindent
We insert the last result in the Dirac equation (3.56). 
Cancelling the exponential factor, it remains the following matrix equation
for the Dirac spinor:

$$[\lambda \epsilon({\bf p})\gamma^0-({\bf p}\gamma) -m]
u(\lambda,{\bf p},\sigma)=0\eqno(3.59)$$
As in eq.(3.43), we write the four component Dirac spinor in terms of two 
two-component ones:
$$u(\lambda,{\bf p},\sigma)= \left[ \matrix{ \varphi \cr
                       \chi }\right]\eqno(3.60)$$
where $\varphi, ~ \chi$ are respectively defined as \textit{upper} and 
\textit{lower} components of the spinor. For brevity we do not write the indices
$\lambda,~{\bf p},~\sigma$ in  $\varphi$ and  $ \chi$.

\noindent
Using the $\gamma^\mu$  in the standard representation of eq.(3.47), 
we can write eq.(3.59) in the form:
$$(\lambda \epsilon({\bf p})-m)\varphi -({\bf p}\sigma)\chi=0  \eqno(3.61a)$$
$$(\lambda \epsilon({\bf p})+m)\chi   -({\bf p}\sigma)\varphi=0\eqno(3.61b)$$
Considering positive energy states, that is $\lambda=+1$, 
we obtain the lower components $\chi_+$ in terms of $\varphi_+$ by means of
eq.(3.61b):
$$\chi_+=   {{({\bf p}\sigma)}\over{ \epsilon({\bf p})+m}} ~\varphi_+\eqno(3.62a)$$
In this case
it is not possible to write $\varphi_+$ in terms of $\chi_+$ 
using eq.(3.61a) because, with $\lambda=+1$,
the factor $\lambda \epsilon({\bf p})-m$ is vanishing for ${\bf p}=0$.

\noindent
Conversely, for negative energy states, that is $\lambda=-1$, from eq.(3.61a)
we obtain the upper components:
$$\varphi_-=  - {{({\bf p}\sigma)}\over{ \epsilon({\bf p})+m}} ~\chi_-\eqno(3.62b)$$
In this way we have found the plane wave solutions of Dirac equation for 
a noninteracting particle. 
The two-component spinors $~\varphi_+ ~, ~~\chi_- $ can be chosen (but it is not
the only possible choice), as those of the nonrelativistic theory.
Denoting them as $w_\sigma$,
with the property $~w^\dagger_{\sigma'} w_\sigma =\delta_{\sigma'\sigma} $,
one has explicitly
$$w_+=\left( \matrix{1 \cr 0} \right)~,~~~~~~~~~~
 w_-=\left( \matrix{0 \cr 1} \right)$$
for spin \textit{up} and \textit{down}, respectively.

\noindent
In consequence
the Dirac spinors $u(\lambda,{\bf p},\sigma)$ can be put in the form

$$u(+1,{\bf p},\sigma)=N \left[ \matrix{ w_\sigma \cr
     {{({\bf p}\sigma)}\over{ \epsilon({\bf p})+m}}~  w_\sigma }\right]\eqno(3.63a)$$
and

$$u(-1,{\bf p},\sigma)=N \left[ \matrix{ -{{({\bf p}\sigma)}\over{ \epsilon({\bf p})+m}}  
~w_\sigma \cr
     w_\sigma }\right]\eqno(3.63b)$$
We point out that, in general, the spin label $\sigma$ of $w_\sigma$ 
\textit{ does not represent}
the spin eigenvalue in a fixed direction, for example the $x^3$ axis. 
This property holds true 
\textit{only} for a particle at rest. 
In this case the previous solutions coincide with the solutions of eq.(3.44).

\noindent
General properties of spin and angular momentum for Dirac equation 
will be studied in a subsequent work.

\vskip 0.5 truecm
\noindent
The  Dirac spinors of eqs.(3.63a,b)  can be also conveniently written as
$$u(\lambda,{\bf p},\sigma)=N~u(\lambda,{\bf p})w_\sigma \eqno(3.64)$$
with 
$$u(+1,{\bf p})=N \left[ \matrix{ 1 \cr
     {{({\bf p}\sigma)}\over{ \epsilon({\bf p})+m}} }\right]\eqno(3.65a)$$
and
$$u(-1,{\bf p})=N \left[ \matrix{ -{{({\bf p}\sigma)}\over{ \epsilon({\bf p})+m}}  
 \cr
     1 }\right]\eqno(3.65b)$$
where the $u(\lambda,{\bf p})$ represent  $4\times 2$ matrices. 
They must be applied onto the
two-component (column) spinors $w_\sigma$, giving as result the four component
Dirac (column) spinors of eqs.(3.63a,b).

\noindent
Note that, in contrast to the nonrelativistic case, the Dirac spinors \textit{depend}
on the momentum  of the particle.

\vskip 0.5 truecm
\noindent
We now discuss the normalization factor $N$.
In nonrelativistic theory, the plane wave of a spin $1/2$ particle is ``normalized''
as 
$$\psi_{{\bf p} \sigma}(x)={1\over\sqrt V} w_\sigma 
\exp \left[i (- E t +{\bf p r})\right]$$
where $V$ represents the  (macroscopic) volume where the particle stays.
The probabilty of finding the particle in this volume is set equal to one.
However, $V$ is a \textit{fictitious} quantity that always disappears when
physical (observable) quantities are calculated. 
In consequence, for the sake of simplicity, one can put $V=1$.
In this way, one has 
$$\psi^\dagger_{{\bf p} {\sigma'}}(x) \psi_{{\bf p} \sigma}(x)= 
\delta_{\sigma {\sigma '}}$$
A similar result can be obtained for the Dirac equation plane waves, putting
in eqs.(3.63a)-(3.65b)
$$N=N^{nc}=\sqrt{{\epsilon({\bf p})+m}\over {2 \epsilon({\bf p})}} \eqno(3.66)$$
where $nc$ stands for \textit{not covariant }. In fact this normalization
cannot be directly used for the calculation of covariant amplitudes. 
With this noncovariant normalization,
the Dirac wave function satifies the following normalization equation
that is analogous to the nonrelativistic one

$$\psi^\dagger_{{\lambda'} {\bf p} {\sigma'}}(x) 
\psi_{{\lambda} {\bf p} \sigma}(x)= 
\delta_{\lambda {\lambda'}} \delta_{\sigma {\sigma '}}\eqno(3.67)$$

\vskip 0.5 truecm
\noindent
As an exercise, verify this result and that of eq.(3.69),  by using eq.(A.7) 
for the products of $(\sigma {\bf p})$. 
Also use the identity
$${\bf p}^2= [\epsilon({\bf p})]^2 - m^2= 
(\epsilon({\bf p})+m)(\epsilon({\bf p})-m)$$

\vskip 0.5 truecm
\noindent
The \textit{covariant normalization} is obtained taking

$$N=N^{cov}=\sqrt{{\epsilon({\bf p})+m}\over {2 m}}=
\sqrt{{\epsilon({\bf p})}\over { m}}~N^{nc} \eqno(3.68)$$
By using this normalization one has
$$\bar u({\lambda '}, {\bf p}, {\sigma '}) u(\lambda, {\bf p}, \sigma)
=(-1)^\lambda \delta_{\lambda {\lambda'}} \delta_{\sigma {\sigma'}}
\eqno(3.69)$$
that, also recalling eq.(3.52a), represents an explicitly Lorentz invariant condition.

\vskip 0.5 truecm
\noindent
In many textbooks a slightly different covariant normalization in used,
that is
$$N^{cov '}= N^{cov}\sqrt{2m}$$
so that a factor $2m$ appears in the \textit{r.h.s.} of eq.(3.69).

\noindent
When reading a book or an article for the study of a specific problem,
pay attention to which normalization is really used !

\vskip 0.5 truecm
\noindent
For further developments it is also introduced the spinor corresponding to
negative energy, \textit{negative momentum} ${\bf -p}$ (and spin label $\sigma$).
From eq.(3.63b) or (3.65b) one has
$$u(-1,{\bf- p})=N \left[ \matrix{ {{({\bf p}\sigma)}\over{ \epsilon({\bf p})+m}}  
 \cr
     1 }\right]\eqno(3.70)$$
Note that 
$$u(-1,{\bf- p})=-\gamma^5 u(+1,{\bf p}) \eqno(3.71)$$
That spinor is standardly applied to $w_\sigma$, as in eqs.(3.63b) and (3.64).

\vskip 0.5 truecm
\noindent
We conclude this section studying the  transition current associated to 
the Dirac equation in the same way as we studied that of  the Klein-Gordon equation
in eqs.(3.6)-(3.8b).

\noindent
First, one has to write the Dirac equation for the \textit{adjoint} wave function
$$\bar \psi(x)= \psi^\dagger(x)\gamma^0$$
To this aim, take the Dirac equation (3.56) and calculate the hermitic conjugate.
By using eq.(3.39), one finds

$$-i\partial_\mu \psi^\dagger(x)\gamma^0 \gamma^\mu \gamma^0 
-\psi^\dagger(x)m =0 \eqno(3.72a)$$
Multiplying this equation from the right by $-\gamma^0$ one obtains
$$i\partial_\mu \bar \psi(x)\gamma^\mu 
+\bar\psi(x) m =0 \eqno(3.72b)$$
that is the searched equation.

\noindent
As done for the Klein-Gordon equation we obtain the conserved current by means of
the following three steps.

\noindent
(i) Take eq.(3.56) with a plane wave, initial state,  solution $\psi_I(x)$ 
corresponding to
energy sign $\lambda_I$, three-momentum ${\bf p}_I$ and spin  label $\sigma_I$.

\noindent
(ii) Analogously, take eq.(3.72b) with a plane wave, final state, solution
  $\bar \psi_F(x)$. 

\noindent
 (iii) Multiply the equation of step (i) by $\bar \psi_F(x)$ and 
the equation of
 
\noindent
step (ii) by  $\psi_I(x)$. Then \textit{sum} these 
two equations (note that the scalar mass term disappears), obtaining
$$\partial_\mu J^\mu_{FI}(x)=0\eqno(3.73)$$
where the Dirac \textit{conserved current} is

$$ J^\mu_{FI}(x)=\bar \psi_F(x) \gamma^\mu \psi_I(x)
\eqno(3.74a) $$
$$=\bar u(\lambda_F,{\bf p}_F,\sigma_F) \gamma^\mu
u(\lambda_I,{\bf p}_I,\sigma_I) \exp(i q_\mu x^\mu) \eqno(3.74b)$$
with the four-momentum tranfer $q^\mu=p_F^\mu-p_I^\mu$.
The four-vector character of the Dirac current is manifestly shown by the
previous equation. 

\noindent
The Dirac four-vector vertex is
$$\bar u_F\gamma^\mu u_I= \bar u(\lambda_F,{\bf p}_F,\sigma_F) \gamma^\mu
u(\lambda_I,{\bf p}_I,\sigma_I)\eqno(3.75)$$
Due to current conservation it satisfies, analogously to eq.(3.9),
$$q_\mu \bar u_F\gamma^\mu u_I=0\eqno(3.76)$$
Note that in the static case the current density (differently from
the Klein Gordon equation) is a positive quantity
both for positive and negative enery states,
as shown explictly by the second equality of the following equation:
$$J^0_{I I}=\bar\psi_I(x)\gamma^0 \psi_I(x)=
\psi^\dagger_I(x) \psi_I(x) > 0 \eqno(3.77)$$
This property allows to attach (for some specific problems) 
a probabilistic interpretation to that quantity
and to consider $\psi(x)$ as a wave function in the same sense of nonrelativistic
quantum mechanics. 
However, the presence of negative energy solutions requires, in general, 
the introduction of the field theory formalism.

\vskip 0.5 truecm
\noindent
The vertex of eq.(3.75) at first glance looks very different with respect to that of
the Klein-Gordon equation $(p^\mu_F+ p^\mu_I)$ given in eq.(3.8b).
The so-called Gordon decomposition, with some algebra on the Dirac matrices,
shows that it can be written in a form that is more similar to the Klein-Gordon one.
This procedure will be analyzed in a subsequent work.

\noindent
For the moment, using the properties of the Pauli matrices, the reader can show that
$$\bar u(\lambda,{\bf p},\sigma ') \gamma^\mu
u(\lambda,{\bf p},\sigma)={p^\mu \over m}\delta_{\sigma \sigma '}
\eqno(3.78)$$
with $p^\mu=(\epsilon({\bf p}), {\bf p})$. The covariant normalization of eq.(3.68) 
has been used.

\vskip 0.5 truecm
\noindent
We conclude this work noting that, at this point, the reader should be able to use 
the main tools related to Dirac equation, 
being also familiarized with the issues of relativity in
quantum mechanical theories.

\noindent
More formal details and calculations of physical observables can be found in many
textbooks and will be studied in a subsequent work.

\section{Appendix. Properties of the Pauli Matrices }

\vskip 0.5 truecm
\noindent
The three Pauli matrices are defined as follows
$$\sigma^1=    \left[ \matrix{ 0 ~~~~~~  1 \cr
                               1 ~~~~~~  0}\right],~~~~~
  \sigma^2=    \left[ \matrix{ 0 ~~~  -i \cr
                               i ~~~~~~  0}\right],~~~~~
  \sigma^3=    \left[ \matrix{1 ~~~~~~  0 \cr
                               0 ~~~-1}\right]\eqno(A.1)$$
they are $2 \times 2$, traceless, hermitic 
(${\sigma^\alpha}^\dagger=\sigma^\alpha$) matrices.
The Pauli matrices fulfill the the following commutation rules  
$$[\sigma^\alpha,\sigma^\beta]=2i\epsilon^{\alpha \beta \gamma}\sigma^\gamma \eqno(A.2)$$
One defines the \textit{spin}, that is the intrinsic angular momentum operator,
multiplying the $\sigma^\alpha$ by $\hbar/2$.

\noindent 
By means of this definition, the spin satisfies the standard angular 
momentum commutation rules, that are

$$[j^\alpha, j^\beta]=
i\hbar \epsilon^{\alpha \beta \gamma} j^\gamma$$ 
\textit{Independently}, the Pauli matrices fulfill the anticommutation rules
$$\{ \sigma^\alpha, \sigma^\beta \}=2 \delta^{\alpha \beta} \eqno(A.3)$$ 
Summing up eqs.(A.2) and (A.3) and dividing by two, one obtains the very useful relation
$$\sigma^\alpha \sigma^\beta= \delta^{\alpha \beta} + 
i\epsilon^{\alpha \beta \gamma}\sigma^\gamma \eqno(A.4)$$
Obviously \textit{only two} of eqs.(A.2), (A.3) and (A.4) are independent.

\vskip 0.5 truecm
\noindent
Given the three-vectors ${\bf a}$ and ${\bf b}$,
let us multiply the previous expression by  $a^\alpha$ and
$b^\beta$, summing over the components. One obtains
$$(\sigma {\bf a}) (\sigma {\bf b})= {\bf a b}
+i(\sigma {\bf a}\times {\bf b})\eqno(A.5)$$
Note that $(\sigma {\bf a})$ represents the following matrix
$$(\sigma {\bf a})=\left[ \matrix{a^3 ~~~~~~ a^1-i a^2 \cr
                                a^1+i a^2 ~~~-a^3}\right]\eqno(A.6)$$
and analogously for $(\sigma {\bf b})$ and $(\sigma {\bf a}\times {\bf b})$.

\noindent 
In eq.(A.5), if ${\bf b}={\bf a}$, the vector product is vanishing, so that one has
$$(\sigma {\bf a})^2= {\bf a}^2 \eqno(A.7)$$
Starting from this equality we can calculate the \textit{function}
$f(\sigma {\bf a})$.

\noindent
To this aim we recall that, if a function $f(x)$ of a standard variable $x$ has 
the Taylor expansion
$$f(x)=\sum_{n=0}^\infty c_n x^n \eqno(A.8)$$
the (same) function of the \textit{matrix} $(\sigma {\bf a})$ 
is \textit{defined} as follows
$$f(\sigma {\bf a})=\sum_{n=0}^\infty c_n   (\sigma {\bf a}) ^n \eqno(A.9)$$
The result is obviously a $2\times 2$ matrix. 

\noindent
Incidentally,
the previous definition, that makes use of the Taylor expansion  in powers of the
\textit{argument} matrix, 
is a general one: it holds  not only for $ (\sigma {\bf a})$ but also
if the argument of the function is a matrix 
of \textit{any} dimension or if it is a linear operator. 
In the present case,
the powers $(\sigma {\bf a}) ^n $ in eq.(A.9) can be calculated by means of eq.(A.7).
We also use $(\sigma {\bf a}) ^0 =1 $.

\noindent 
We make here some algebraic developments to obtain  a ``closed'' expression
for eq.(A.9).

\noindent
First, let us write separately the even and the odd powers in the expansion (A.8):
$$f(x)=\sum_{m=0}^\infty c_{2m} x^{2m} +
\sum_{l=0}^\infty c_{2l+1} x^{2l+1}
 \eqno(A.10)$$
Do the same for $f(-x)$:
$$f(-x)=\sum_{m=0}^\infty c_{2m} x^{2m} -
\sum_{l=0}^\infty c_{2l+1} x^{2l+1}
 \eqno(A.11)$$
So that, summing and subtracting the last two equations, one has:

$${1\over 2}[f(x)+f(-x)]=\sum_{m=0}^\infty c_{2m} x^{2m} \eqno(A.12)$$
$${1\over 2}[f(x)-f(-x)]=\sum_{l=0}^\infty c_{2l+1} x^{2l+1} \eqno(A.13)$$
Let us now go back to eq.(A.9), introducing the unit vector ${\bf \hat a}$
and the absolute value (\textit{positive} !) $|{\bf a}|$,
by means of the standard relation

$${\bf a}=|{\bf a}|{\bf \hat a} \eqno(A.14)$$
Furthermore, by means of eq.(A.7), one has
$$(\sigma {\bf a}) ^{2m}=({\bf a}^2)^m=|{\bf a}|^{2m} \eqno(A.15)$$
$$(\sigma {\bf a}) ^{2l+1}=   (\sigma {\bf a})({\bf a}^2)^l =
 (\sigma {\bf \hat a})|{\bf a}|^{2l+1} \eqno(A.16)   $$
In consequence, writing separately the even and odd powers in eq.(A.9),
and using eqs.(A.12,13), one obtains
$$f(\sigma {\bf a})=\sum_{m=0}^\infty c_{2m}|{\bf a}|^{2m}+
(\sigma {\bf \hat a})\sum_{l=0}^\infty c_{2l+1}|{\bf a}| ^{2l+1}=$$
$$= {1\over 2}[f(|{\bf a}|)+f(-|{\bf a}|)] +
 {1\over 2}[f(|{\bf a}|)-f(-|{\bf a}|)] (\sigma {\bf \hat a}) \eqno(A.17)$$

\vskip 0.5 truecm
\noindent
In order to derive the second equality of eq.(3.17b), being the $\alpha^\delta$,
defined in eq.(3.11) as
block \textit{diagonal} matrices, one can procede separately for the two blocks.
Let us consider first the upper left block.
By means of the previous equation, one has
$$\exp\left[-{\omega\over 2}(\sigma {\bf \hat v}) \right]=$$
$$ ={1\over 2}\left[\exp(|{\omega \over 2}|) +\exp(-|{\omega \over 2}|)\right]
   -{1\over 2}\left[\exp(|{\omega \over 2}|) -\exp(-|{\omega \over 2}|)\right]
sgn(\omega)(\sigma {\bf \hat v})=$$
$$=ch({\omega \over 2}) -   (\sigma {\bf \hat v}) sh({\omega \over 2})\eqno(A.18)$$
In the previous equation $sgn(\omega)$ gives the sign of $\omega$. 
Also, we have used
${\bf \hat a}=-sgn(\omega){\bf \hat v}$ and $|{\bf a}|=|\omega|$
in eq.(A.17).

\noindent
As for  the lower right block, one easily obtain the result that is analogous 
to the previous one, but with a plus sign in front of the second term.
Recalling the form of the $\alpha^\delta$ matrices, one obtains the final result
of eq.(3.17b).

\vskip 0.5 truecm
\noindent
The reader can now look at this development in a slightly different way.
Recalling the form of the $\alpha^\delta$ of eq.(3.11),
the powers of $(\alpha {\bf a})$,  satisfy the \textit{same} relations as
eqs.(A.15) and (A.16) for the powers of $(\sigma {\bf a})$.
In consequence, one can repeat the calculations of eqs.(A.17) and (A.18) simply
replacing the $\sigma^\delta$ with the $\alpha^\delta$, obtaining directly 
eq.(3.17b).

\noindent
Furthermore, in this way, one  realizes that the result remains the same also
in the standard representation and only depends on the anticommutation rules
of the $\alpha^\delta$ matrices.


\end{document}